\documentclass[notitlepage,twocolumn,letterpaper,natbib,aps,prl,amsmath,amsfonts,nofootinbib,preprintnumbers,superscriptaddress,secnumarabic]{revtex4-1}
\pdfoutput=1
\usepackage{amssymb,amsmath,latexsym,mathrsfs}
\usepackage{url}
\usepackage{enumitem}
\usepackage{bm}
\usepackage[normalem]{ulem}
\usepackage{graphicx}
\usepackage{booktabs}
\usepackage[usenames,dvipsnames]{color}
\usepackage[breaklinks,colorlinks,urlcolor=magenta,citecolor=magenta,linkcolor=magenta]{hyperref}
\usepackage{multirow}
\usepackage{float}
\usepackage{cases}
\usepackage{blindtext}
\usepackage{pifont}
\usepackage{hhline}
\usepackage{siunitx}
\usepackage{braket}
\usepackage{aas_macros}
\usepackage{bbold}
\usepackage{fontawesome}
\usepackage[table,xcdraw]{xcolor}
\definecolor{linkcolor}{rgb}{0.0, 0.47, 0.75}
\definecolor{citecolor}{rgb}{1.0, 0.5, 0.0}
\definecolor{linkcolor}{rgb}{0.390625,0.5607843137,0.99609375}
\hypersetup{
  linkcolor  = linkcolor,
  citecolor  = linkcolor,
  urlcolor   = linkcolor,
  colorlinks = true
}
\hypersetup{
  linkcolor  = linkcolor,
  citecolor  = linkcolor,
  urlcolor   = linkcolor,
  colorlinks = true
}
\makeatletter
\def\maketitle{
\@author@finish
\title@column\titleblock@produce
\suppressfloats[t]}
\makeatother

\usepackage{comment}
\begin{document}

\title{%
    Dwarf galaxies imply dark matter is heavier than
    \texorpdfstring{$2.2 \times 10^{-21} \, \text{eV}$}{2.2 zeV}
}

\author{Tim Zimmermann}
\email{timzi@astro.uio.no}
\affiliation{Institute of Theoretical Astrophysics, University of Oslo, P.O. Box 1029 Blindern, Oslo, Norway}

\author{James Alvey}
\email{jbga2@cam.ac.uk}
\affiliation{Kavli Institute for Cosmology Cambridge, Madingley Road, Cambridge CB3 0HA, United Kingdom}
\affiliation{Institute of Astronomy, University of Cambridge, Madingley Road, Cambridge CB3 0HA, United Kingdom}
\affiliation{GRAPPA Institute, Institute for Theoretical Physics Amsterdam, University of Amsterdam, Science Park 904, 1098 XH Amsterdam, The Netherlands}

\author{David J. E. Marsh}
\email{david.j.marsh@kcl.ac.uk}
\affiliation{Theoretical Particle Physics and Cosmology, King’s College London, Strand, London, WC2R 2LS, United Kingdom}

\author{Malcolm Fairbairn}
\email{malcolm.fairbairn@kcl.ac.uk}
\affiliation{Theoretical Particle Physics and Cosmology, King’s College London, Strand, London, WC2R 2LS, United Kingdom}

\author{Justin I. Read}
\email{j.read@surrey.ac.uk}
\affiliation{Department of Physics, University of Surrey, Guildford, GU2 7XH, UK}

\preprint{KCL-TH-PH-2024-30}

\begin{abstract}

\noindent %
It is widely established that a lower bound on the dark matter 
particle mass, $m$,
can be obtained by demanding that the de Broglie wavelength in a given galaxy must be 
smaller than the virial radius of the galaxy, leading to $m\gtrsim 10^{-22}\text{ eV}$ 
when applied to typical dwarf galaxies. 
This lower limit has never been derived precisely or rigorously. We use stellar kinematical 
data for the Milky Way satellite galaxy Leo II to self-consistently reconstruct a statistical 
ensemble of dark matter wavefunctions and corresponding density profiles. 
By comparison to a data-driven, model-independent reconstruction, and using a variant of the 
maximum mean discrepancy as a statistical measure, we determine that a self-consistent 
description of dark matter in the local Universe requires 
$m>\SI{2.2e-21}{\electronvolt}\;\mathrm{(CL>95\%)}$. 
This lower limit is free of any assumptions pertaining to cosmology, microphysics 
(including spin), or dynamics of dark matter, and only assumes that it is predominantly 
composed of a single bosonic particle species.

\vspace*{12pt} 
\noindent \textbf{\texttt{GitHub}}: The \textsc{jaxsp} library is available at 
\href{https://github.com/timzimm/jaxsp}{\faGithub\;timzimm/jaxsp}. 
In addition, the scripts to generate the results in this work can be found at 
\href{https://github.com/timzimm/boson_dsph}{\faGithub\;timzimm/boson\_dsph}.

\vspace*{32pt}
\end{abstract}

\maketitle
\hypersetup{
  linkcolor  = linkcolor,
  citecolor  = linkcolor,
  urlcolor   = linkcolor
}

\renewcommand*{\thefootnote}{\arabic{footnote}}
\setcounter{footnote}{0}

\noindent \textbf{Introduction.} Astrophysical evidence for a large amount of gravitationally interacting matter, which cannot be explained in the context of the Standard Model of particle physics and General Relativity, has been accumulating for more than 100 years~\cite{Bertone:2016nfn,Marsh:2024ury}. Nonetheless, the identity and precise nature of dark matter (DM) remains one of the biggest questions in our understanding of the Universe. There is a myriad of evidence – coming from, e.g., observations of the cosmic microwave background~\cite{Planck:2018vyg}, galaxy clusters (e.g.~\cite{BOSS:2016wmc}), large-scale structure (e.g.~\cite{SDSS:2005xqv}), or stellar kinematics (e.g.~\cite{Battaglia:2008jz}) – for the existence of such a non-baryonic matter component. To date, however, no definitive laboratory or non-gravitational signature has been observed. As such, despite the efforts of direct detection facilities (e.g.~\cite{XENON:2018voc, ADMX:2021nhd}), large-scale cosmology surveys~\cite{Planck:2018vyg,SDSS:2005xqv,BOSS:2016wmc}, and telescopes (see Ref.~\cite{Gaskins:2016cha} for a review of indirect detection searches), the dark matter parameter space remains wide open.

If dark matter is thought of as dominantly composed of a single particle or composite object, it is natural to categorise DM according to the mass, $m$, of the constituent. At the highest end, we have macroscopic objects, such as black holes or other massive objects. Dynamical effects of these objects on star clusters, for example, lead to an upper limit on the allowed mass in the range of 10 solar masses ($M_\odot$)~\cite{Brandt:2016aco}. For $m<M_{\rm Pl}$, where $M_{\rm Pl}$ is the Planck mass, DM may be composed of a new fundamental particle. For decades, experimental programs have been developed to search for such particles, for example scattering of DM with $m\approx 1\text{ GeV}$ off atomic nuclei (e.g.~\cite{XENON:2018voc}), or resonant production of radio waves in microwave cavities for DM with $m\approx 1 \,\mu\text{eV}$ (e.g.~\cite{ADMX:2021nhd}). 

In this \emph{Letter} we ask the simple question: what is the smallest value $m$ can possibly take, consistent with kinematical considerations? Specifically, we are interested in the implications of kinematical observations of Milky Way dwarf spheroidal galaxies.

If DM is a fermion, then the resulting lower limit is known as the Tremaine-Gunn bound~\cite{Tremaine:1979we}. As a result of the Pauli exclusion principle, fermions cannot multiply occupy states in phase space, and so bounding the distribution function $f\leq 1$, one can estimate a lower limit on $m$ given the mass and radius (and hence virial equilibrium velocity) of a galaxy. Modelling a dwarf galaxy as a homogeneous, spherically symmetric and  degenerate Fermi gas with $M\simeq10^8 \, M_\odot$ and $R\simeq 1.0\text{ kpc}$, this gives an estimated lower bound $m_{\rm f}\gtrsim 120 \text{ eV}$~\cite{Boyarsky2009}. This limit can be improved by accurately reconstructing the phase space distribution function via the observed kinematics of tracer stars hosted by the Milky way dwarfs, which leads to $m_{\rm f}>130 \text{ eV}$ at the 95\% confidence level (CL)~\cite{Alvey:2020xsk}. However, this limit does not apply if DM is a boson.

For a boson, a limit as fundamental as the fermionic Tremaine-Gunn bound can be estimated by invoking the uncertainty principle $m\sigma_v\sigma_r \gtrsim 3/2\times\hbar$~\citep{Rudnicki2012}. For a steady-state, spherically symmetric object like a dwarf galaxy composed of bosonic DM to exist, its spatial extent $r_{\rm vir}$ and characteristic velocity dispersion $\sigma_v$ must satisfy $m \gtrsim 3/2 \times \hbar \left(\sigma_v r_{\rm vir}\right)^{-1}$. If we estimate the virial radius, $r_{\rm vir}$, and velocity dispersion, $\sigma_v$, for the Milky Way dwarf Leo II \citep{Spencer2017} as $r_{\rm vir}\simeq 9\;\mathrm{kpc} $ and $\sigma_v\simeq15\;\mathrm{km\;s^{-1}}$~\citep{Read2017, Collins2021},\footnote{As is customary, we set $r_{\rm vir} = r_{200}$, i.e. the radius within which the mean density equals $200$ times the critical density, and estimate $\sigma_v$ via the scalar virial theorem \citep{Binney1987}: $\sigma_v^2~=~2\pi M_{\rm tot}^{-1}\left|\int\text{d}rr^2 \rho(r)V(r)\right|$.} this leads to 
$m \gtrsim \SI{2e-23}{\electronvolt}$. This is at the same order of magnitude as the
commonly assumed lower limit on the DM particle mass.
In the rest of this \emph{Letter}, we work to derive a more
rigorous limit, which will turn out to be two orders of magnitude stronger. In particular, we strengthen the limit to $m > 2.2 \times 10^{-21}\,\mathrm{eV}\;\mathrm{(CL>95\%)}$. Of course, there are a number of other relevant limits on ultra-light dark matter candidates, see in particular Refs.~\cite{Rogers:2020ltq,Marsh2019,Dalal2022}. Such limits, however, depend on details of cosmology, non-linear gas physics, and/or long-time dynamics, all of which our limit is independent of. We postpone a detailed comparison and discussion to the end of this work.

Input to our rigorous analysis is a set of $5000$ stationary and spherically symmetric density-potential pairs derived from stellar kinematical data of Leo II. 
Our mass limit is then derived in two steps: 
For each potential, we reconstruct a DM wave function as an exhaustive expansion of its energy 
eigenstates. 
We then compare the set of reconstructed wave functions and data-driven input densities at the
population level and vary the bosons mass $m$ until both sets are statistically indistinguishable.
We formalise this idea as a non-parametric, two-sample
hypothesis test based on the \emph{fused maximum mean discrepancy}
\citep{Gretton2012, Biggs2023} for functional data \citep{Wynne2020}. The
analysis relies on \textsc{jaxsp}, our differentiable and scalable wave function
reconstruction tool based on a semi-analytical treatment of Schr\"{o}dinger's
equation. We describe \textsc{jaxsp} in the \emph{Supplemental Material} \cite{supp}.

\begin{figure}
    \centering
    \includegraphics[width=\columnwidth]{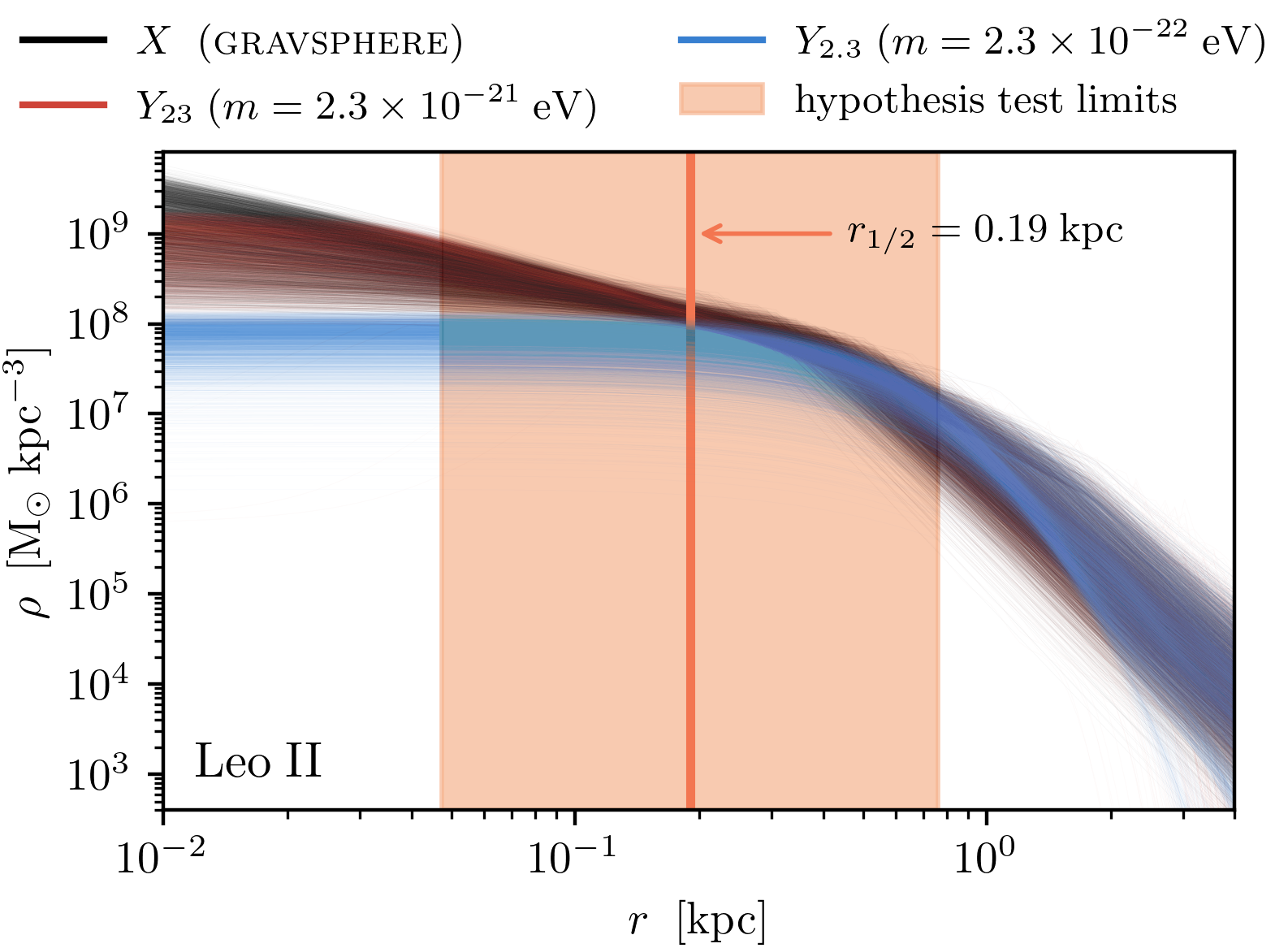}
    \caption{%
        Direct comparison of different density ensembles for the Milky way dwarf Leo II.
        Black curves depict density samples generated by the Jeans code \textsc{gravsphere} 
        and act as data-driven input set $X$ to our analysis pipeline.
        The pipeline output is a population $Y_{m_{22}}$ of reconstructed DM wave functions densities $\langle|\psi|^2\rangle$ at boson mass $m = m_{22}\times10^{-22}\mathrm{eV}$ (red/blue curves).
        Higher mass eigenstates enjoy a stronger spatial localisation, and are therefore able to resolve more
        structure of a cuspy input density at small radii. Testing this small scale discrepancy in the validity region of \textsc{gravsphere} (orange sector) at the population level, cf. Fig. \ref{fig:hypothesis_test}, summarises the key idea behind our lower mass limit.
    }
    \label{fig:gravsphere_posterior}
\end{figure}

\vspace{8pt}
\noindent\textbf{Leo II dSph data.} To carry out the wavefunction fit, we require robust data-driven reconstructions of the DM density profiles within dwarf galaxies. Specifically, we need a statistical ensemble of spherically symmetric DM density profiles $\rho(r)$ that are consistent with the photometric and kinematical measurements of the stellar tracer population in Leo II~\cite{Spencer2017}. For this we use \textsc{gravsphere} \citep{Read2017,
Read2018, Genina2020, Collins2021}, a
Markov-Chain-Monte-Carlo sampler that infers realisations of $\rho$ by solving the spherical Jeans equation 
\citep{Jeans1922, Binney1982, Binney1987}. To mitigate the well-known
degeneracy between the halo density and the stellar velocity anisotropy inherent to Jeans
modelling \citep{Merrifield1990, Wilkinson2002, Lokas2003, Gonzalez2017}, 
\textsc{gravsphere} incorporates higher order velocity moments, so called virial shape parameters
\citep{Merrifield1990,Richardson2014,richardson_analytical_2013}, into its likelihood.
The results of this analysis are posterior parameter samples for the highly flexible,
\textsc{coreNFWtides}-density model, $\rho_\text{cNFWt}$. This density model augments the canonical NFW profile \citep{Navarro1997} with four additional parameters. At small
radii, a `coredness' parameter,  $0\leq n \leq 1$, and core radius $r_c$ permit us to interpolate between a 
perfect
core $\rho(r \ll r_c) \sim \rho_0$ or cusp $\rho(r\ll r_c) \sim r^{-1}$. At large radii, the
introduction of a tidal radius $r_t$ and power law decay $\rho(r\gg r_t) \sim r^{-\delta}$ allow for an 
effective model of tidal forces stripping peripheral mass away, if a larger host halo is
present. This choice to allow for an inner core component accommodates the presence of unmodelled systematics such as e.g. baryonic feedback or wavelike-DM dynamical heating ~\cite{Hui2016, Bar2019}. Since a more cored profile leads to weaker limits within our method, this renders this modelling choice a conservative one. We discuss the impact of additional technical modelling choices, such as the spatial cutoff and sphericity assumptions in the \emph{Supplemental Material}~\cite{supp}.

Our choice of Leo II was motivated by its ability to set a strong limit due to
its steep inner profile, and its statistically robust description as a
static~\cite{Read:2006aaa,DeLeo:2023aaa}, spherically symmetric
distribution~\cite{Read:2017aaa,Read:2018aaa,Genina:2019aaa,DeLeo:2023aaa} with
a well measured cusp. Draco is another possible candidate
object~\cite{Alvarez:2020aaa}, however, it is computationally more challenging
with the current iteration of our method~\cite{supp}. 
We have explicitly checked in a pared-down analysis that Draco gives a consistent limit 
with Leo II. 

Fig.~\ref{fig:gravsphere_posterior} displays the $5000$ posterior samples for
Leo II (black lines) that we use in our analysis.
Note that Collins et al. \citep{Collins2021} also validate \textsc{gravsphere}'s mass modelling on the
spherical mock data described in Ref.~\cite{Read2021} and find it
to recover the true density within the $95\%$ confidence interval of the density
samples over the radial range $0.25 \leq r/r_{1/2} \leq 4$ (orange sector in Fig.~\ref{fig:gravsphere_posterior})
around the half-light radius $r_{1/2}$
even if only $O(100)$ tracers are available. 
We adopt this radial range for our analysis.

\vspace{8pt}
\noindent \textbf{Self-consistent Reconstruction of the DM wave function.} In the non-relativistic limit, the dynamics of spin zero, bosonic DM
with $m\ll\mathcal{O}(10)\mathrm{\;eV}$ are described by a complex scalar $\psi(\bm x, t)$ that obeys the Schr\"odinger-Poisson (SP) equation
\citep{Hui2021} and sources its gravitational potential $V$ via the DM density
$\rho(\bm x, t) = |\psi(\bm x,t)|^2$. Assuming stationary conditions, i.e. no
explicit time dependency in $\rho$ and $V$, and restricting to spherically
symmetric objects, the linear set of equations:
\begin{equation}
    \label{eq:schroedinger}
    \begin{split}
    -\frac{\hbar^2}{2m}\left(\frac{\partial^2}{\partial r^2}
    -\frac{l(l+1)}{r^2}\right)u_{nl} + m V u_{nl} 
    &=E_{nl} u_{nl} \;, \\
    \left(\frac{\partial^2}{\partial r^2} + \frac{1}{r}\frac{\partial}{\partial
    r}\right)V 
    &= 4\pi G \rho \;,
    \end{split}
\end{equation}
with eigenmode $\psi_{nlm}(\bm x,t) = r^{-1}u_{nl}(r)Y^m_l(\phi,
\theta)e^{iE_{nl}t/\hbar}$
has proven to be an effective approximation for the equilibrium phenomena
encapsulated in the full-fledged SP equation \cite{Lin2018, Dalal2021, Yavetz2022}.

From a practical perspective, and following Refs.~\cite{Lin:2018whl,Yavetz2022}, Eq.~\eqref{eq:schroedinger} 
breaks the non-linear relation between wave function $\psi$ and potential $V$. In addition, it approximates $\rho$ as a stationary density that is \emph{independent} of the wave function. 
In this way, Eq.~\eqref{eq:schroedinger} is equivalent to a linear
order perturbative treatment of $\psi$ with $V$ as the zeroth-order potential~\cite{Dalal2021}. 

The full process of this reconstruction of the wave function, \cite{supp}, may be seen as an evolution of the approaches presented in Refs.~\cite{Lin:2018whl,Yavetz2022}. Here we recall the
main three steps: Firstly, given a density realisation, 
we compute its time-independent gravitational potential.
We then compute an extensive library of eigenstates that contains all modes $u_{nl}$
with a classically allowed region within $r_{99}$, i.e. the radius enclosing $99\%$ of
the total mass. This involves repeatedly solving Eq.~\eqref{eq:schroedinger} for
increasing values of the angular momentum quantum number $l$ --- a task we solve in practice by
application of a piecewise, perturbative approximation \cite{supp}.
Lastly, we establish self-consistency between the
density sample $\rho_\text{cNFWt}$ and the steady-state contribution 
$\langle |\psi|^2\rangle = (4\pi r^2)^{-1}\sum_{nl} (2l+1)|a_{nl}|^2 u_{nl}^2(r)$, \cite{Yavetz2022},
by minimising an objective function of our choice over the mode coefficients $|a_{nl}|$ --- in this
work the symmetric Jensen-Shannon divergence.\footnote{Together with the eigenstate library construction and mode coefficient optimisation, for a single representative mass $m = 2.3 \times 10^{-21}\,\mathrm{eV}$, our pipeline takes $\mathcal{O}(2 \, \mathrm{days})$ to run on 100 CPU cores for 5000 profiles. Alongside this, there is typically a significant memory footprint of $\mathcal{O}(2 \, \mathrm{TB})$ when holding the eigenstate libraries in memory during the optimisation step.}

\begin{figure}
    \centering
    \includegraphics[width=\columnwidth]{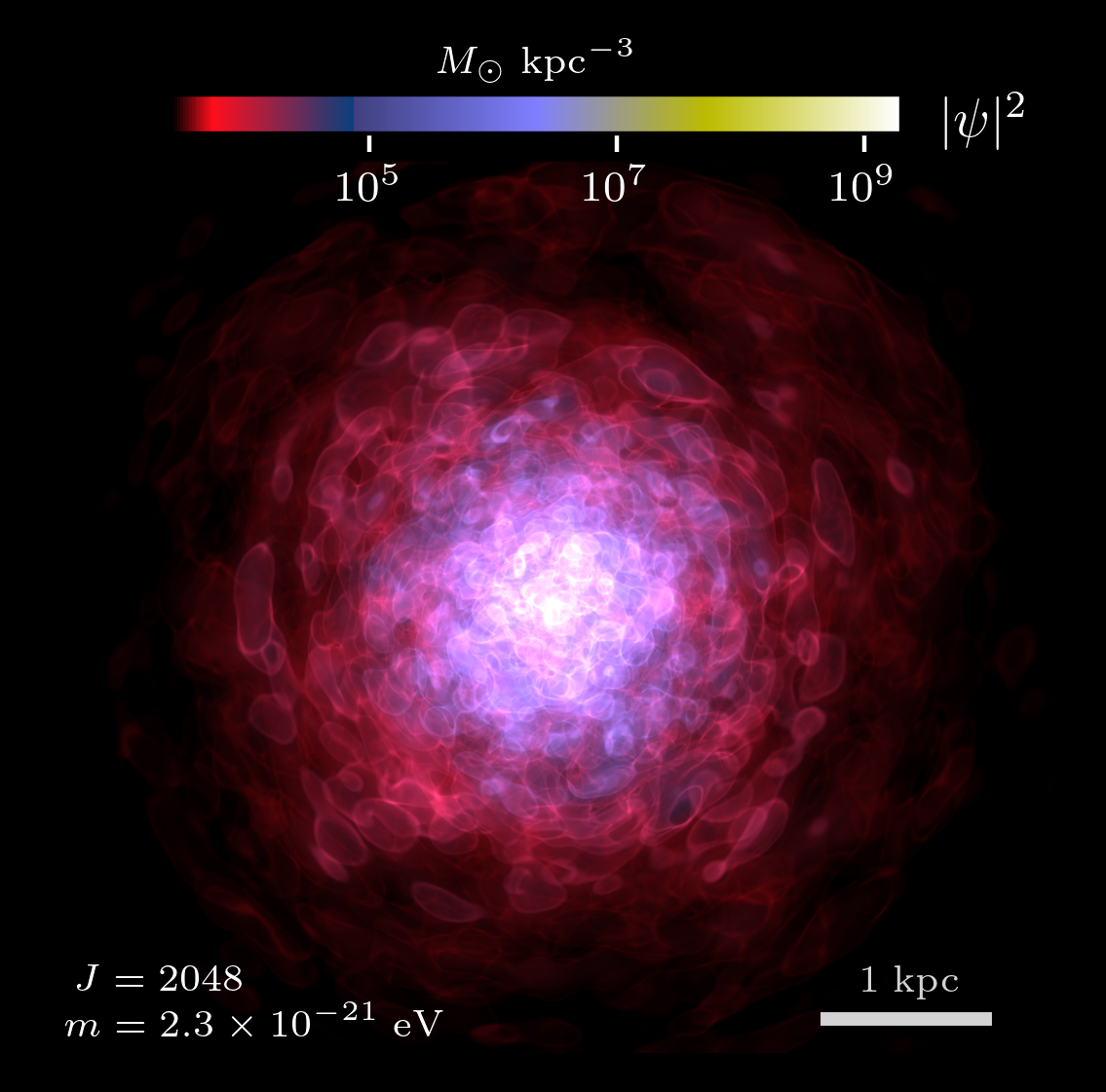}
    \caption{Volume rendering of the total wave function density
    $|\psi|^2$ reconstructed from the average profile $\langle
    \rho_\text{cNFWt}\rangle$ of Fig. \ref{fig:gravsphere_posterior}
    for the allowed boson mass $m = \SI{2.3e-21}{\electronvolt}$
    according to the hypothesis test in Fig. \ref{fig:hypothesis_test}.
    $J$ denotes the total number of radial modes $u_{nl}$ in the eigenstate
    expansion.
    }
    \label{fig:volume_rendering}
\end{figure}

As a concrete example, in Fig.~\ref{fig:gravsphere_posterior} we show the
set of reconstructed $\langle|\psi|^2\rangle$ for 
the input densities $\rho_{\rm cNFWt}$ from \textsc{gravsphere} at two different boson masses -- $m = 2.3 \times 10^{-22} \, \mathrm{eV}$ (excluded by our analysis) and $m = 2.3 \times 10^{-21} \, \mathrm{eV}$ (not excluded). 
In Fig.~\ref{fig:volume_rendering}, we show the full wave function $|\psi(r, \phi, \theta)|^2 = \langle
|\psi|^2\rangle(r) + \chi(r,\phi,\theta)$, including the interference
cross-term $\chi$. The presence of $\chi$ induces intriguing
physics in its own right and may be used to put constraints on the boson mass. Our limit is
derived exclusively from the time independent background
contribution $\langle |\psi|^2\rangle$.

\vspace{8pt}
\noindent \textbf{How to exclude boson masses.} It is perhaps tempting to incorporate an eigenstate expansion approach into
a textbook Jeans analysis, i.e. define a parametrised potential, construct the
corresponding eigenstate library, and incorporate $\langle|\psi|^2\rangle$ into a likelihood to
sample the full posterior distribution of the mode coefficients.
Unfortunately, this is computationally intractable. Depending on the boson mass and depth of the
potential well, eigenstate library sizes of $J=\mathcal{O}(10^4)$ are common,
rendering standard sampling techniques ineffective. 

The question then is how to determine if a given boson mass self-consistently reproduces the true distribution of density profiles as defined by the \textsc{gravsphere} Jeans analysis.
In order to answer this question, we construct a non-parametric, two-sample hypothesis test that 
ascertains whether the set 
of \textsc{gravsphere} functions, $X=\{\rho_{\mathrm{cNFWt},j}\}_{j=1}^{5000}$, and 
the set of wavefunction densities for boson mass $m$, $Y_m=\{\langle|\psi|^2\rangle_j\}_{j=1}^{5000}$, are
drawn from the same underlying distribution. 
Our null hypothesis is therefore $\mathrm{H}_0: p = q_m$, where we denote
$p$ as the ``posterior distribution over density profiles consistent with Leo II stellar 
kinematic data" and $q_m$ as ``the corresponding distribution of reconstructed density profiles 
for a fixed boson mass $m$". 
Our (lower) limit on $m$ at confidence 
level (CL) $(1 - \alpha)$ is then obtained once the null hypothesis cannot be
rejected at the $\alpha$ significance level. Intuitively, this happens at a mass
with eigenstate libraries flexible enough to recover all of the \textsc{gravsphere} densities equally well, irrespective of whether they are cored or cusped.

There are two technical hurdles that we need to overcome to apply this test in practice. Firstly, we need to define a test statistic $\tau(X_1, X_2)$ over two samples of density profiles, $X_{1,2}$. Secondly, at a given boson mass $m$, we need to derive the distribution of $\tau$ under the null hypothesis, $p(\tau \mid \mathrm{H}_0)$, where $X_{1,2}$ are both drawn from the same distribution. With this, we can then evaluate the test statistic on the ``observed" samples $\tau_o = \tau(X, Y_m)$ and compare it to the critical values of the distribution $p(\tau \mid \mathrm{H}_0)$ at the desired significance level.

In this work, we use an estimator for the \emph{fused maximum mean discrepancy}
(MMD) as a test statistic \cite{Sriperumbudur2009, Gretton2012, Wynne2020, Biggs2023}.\footnote{In the context of two-sample, distribution equality testing, \cite{Biggs2023} benchmark the fused maximum mean discrepancy against a selection of state-of-the-art test statistics, and find it to be on par or outperform them in terms correctly rejecting  $\mathrm{H}_0$.
}
We refer to the Supplemental Material, \cite{supp}, for full technical details, however, it can be
shown that MMD constitutes a metric on the space of probability
distributions and that $\tau(X,Y_m)\xrightarrow[]{N\to\infty}\text{MMD}(p, q_m)$, thus
providing an intuitive ``distance-based" interpretation for $\tau$. As such, we can reject $\mathrm{H}_0$ under the observation of sufficiently \emph{large} values of $\tau_o$.

With the test statistic chosen, we follow Ref.~\cite{Biggs2023} and compute $p(\tau |  \mathrm{H}_0)$ by using a permutation
resampling, i.e. we randomly shuffle the combined set $Z=(X,Y_m)$ (mixing
\textsc{gravsphere} and wavefunction densities), split into two equal parts,
$X_1 \cup X_2 = Z$, and compute $\tau(X_1, X_2)$. We repeat this process $\mathcal{O}(10^5)$ times
to obtain an empirical estimate of the cumulative distribution (CDF) of
$p(\tau\mid  \mathrm{H}_0)$ and reject  $\mathrm{H}_0$ if the observed value of $\tau_o$ exceeds the
$(1-\alpha)$ quantile of the CDF 
\footnote{%
    One may show, \citep{Hemerik2018}, that this resampling
    strategy is actually conservative, i.e. constructing the distribution of $\tau$ in this fashion can only weaken the test. 
    We have checked explicitly, however, that using e.g. only the \textsc{gravsphere} functions in the permutation approach 
    we obtain almost identical results.
}.
A more intuitive example of this test applied to the coredness parameter in the \textsc{coreNFWtides} profile is provided in the appendix. 

\begin{figure}  
    \centering
    \includegraphics[width=\columnwidth]{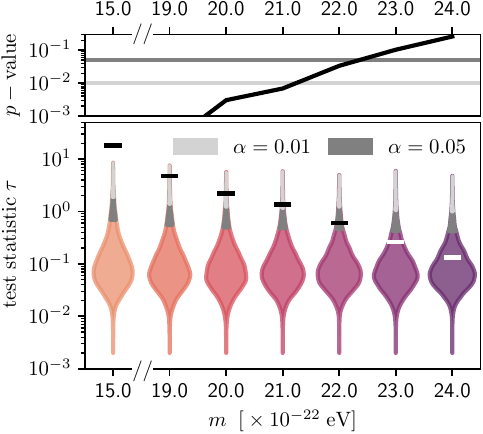}
    \caption{%
        Non-parametric, two-sample hypothesis test for distribution equality based on the fused
        maximum mean discrepancy $\tau(Z)$ 
        as test statistic. Shown is the
        distribution of $\tau(Z)$ under the null hypothesis obtained by a
        permutation resampling. Areas in shades of gray illustrate the rejection regions at significance 
        $\alpha$. Black/white bars depict the realised value of the test statistic.
        We determine $m>\SI{2.2e-21}{\electronvolt}$ as the first mass compatible with the null hypothesis at $\mathrm{CL}>95\%$.
    }
    \label{fig:hypothesis_test}
\end{figure}

\vspace{8pt}
\noindent \textbf{Results.}
Fig.~\ref{fig:hypothesis_test} depicts the distribution of the test statistic
$\tau$ under the null hypothesis $\mathrm{H}_0$, and the
critical values of this distribution at the $\alpha = 1\%, 5\%$ levels. For a
set of boson masses $m \in [15, 24] \times 10^{-22}\,\mathrm{eV}$, we evaluate
the test statistic $\tau_o$ (shown by the black/white horizontal lines in the figure) 
and compare them to the critical values of the distribution.  
As expected, we see that at larger masses, $\tau_o$ is fully consistent with the 
distribution under $\mathrm{H}_0$. In this regime, the reconstructed
wavefunctions are statistically indistinguishable from the \textsc{gravsphere} density profiles. 
On the other hand, at low masses, the eigenstate superposition is unable to
reproduce key features (in particular the cuspier profiles) across the samples,
translating into a statistical discrepancy and rejection of $\mathrm{H}_0$. 

The key result of this work is that we find $m>\SI{2.2e-21}{\electronvolt}\;(\mathrm{CL}>95\%)$. This result is
largely unaffected by moderate changes of the radial range on which the hypothesis test is conducted. Also, evaluating the two-body relaxation rate at this mass gives a timescale much longer than the age of the Universe~\cite{Hui:2016ltb}, making it fully consistent with our modeling of Leo II as stationary.

Our robust lower limit on the DM mass is two orders of magnitude stronger compared to a naive estimate based on the uncertainty principle. The reported improvement may be intuitively understood as a consequence of applying the
uncertainty condition at the level of individual eigenstates rather than at the
level of the overall wave function: Each eigenstate, $\psi_j$, has a spatial extent
set by its uncertainty $\sigma_r^2 = \braket{\psi_j|r^2|\psi_j} -
\braket{\psi_j|r|\psi_j}^2$ and thus contributes to the dwarf density
at that particular length scale. The inferred Leo II DM density is cuspy, i.e. well modeled by a steep power law, on certain length scales. If the length scale of this cusp is comparable to the position uncertainty $\sigma_r$ of the wavefunction, then the eigenstate expansion will not be able to adequately reproduce the density profile. To compensate, the spatial extent of all modes must shrink, requiring an increased boson mass $m$, and tightening the lower limit. If the dark matter distribution would be cored, a lower mass would suffice to reproduce the density ensemble at small radii. We checked this explicitly by applying our analysis to Fornax \cite{Walker2009fornax} yielding a weakened limit of $m>\SI{4e-22}{\electronvolt}$ (broadly consistent with related studies~\cite{Marsh:2015xka,Gonzalez2017}).

\vspace{8pt}
\noindent \textbf{Significance and generality of the result.} There are, of course, many other lower limits on the DM mass that rely on similar physics to that which has been used in this work. We mention a few here for comparison. If the DM relic density arises from coherent motion in a harmonic potential, then the cosmic microwave background anisotropies constrain $m>10^{-24}\text{ eV}$~\cite{Hlozek:2014lca}, which is improved to $m>10^{-23}\text{ eV}$ with weak gravitational lensing~\cite{Dentler:2021zij} and to $m>2\times 10^{-20}\text{ eV}$ including modeling of intergalactic gas in the Lyman-alpha forest~\cite{Rogers:2020ltq}. The same cosmological models can be used to make predictions about galaxy number counts, e.g. Ref.~\cite{DES:2020fxi} finds $m>2.9\times10^{-21}\text{ eV}$ in order to produce the abundance of satellite galaxies of the Milky Way, while Ref.~\cite{Winch:2024mrt} finds $m>2.5\times 10^{-22}\text{ eV}$ to explain high redshift galaxies observed by Hubble and James Webb space telescopes. All of these lower limits are less generic than ours since they rely on assumptions about cosmology and/or complex non-linear dynamics. 

In addition, the physics of the SP system gives rise to a rich 
phenomenology of nonlinear wave effects~\cite{Schive2014}. Dynamical heating of the stellar tracer population~\cite{Hui:2016ltb} in in Eridanus-II and Segue-I and II leads to the lower limit $m>3\times 10^{-19}\text{ eV}$~\cite{Marsh2019, Dalal2022}. This limit is less fundamental than ours, since it relies on dynamics over billions of years (it is dynamical, rather than kinematical), and could be strongly affected by unmodeled events in the history of the galaxy.

We computed our lower limit for a boson with spin $s=0$. Higher spin bosonic DM
is also of interest (see e.g. Ref.~\cite{Jain:2021pnk}). In such a case, there are copies of the DM wave function for each polarisation
state. Taking the superposition of such terms to form the total density reduces the amplitude of the interference term $\chi(r, \phi, \theta)$, but
does not affect the spherically averaged density $\langle|\psi^2|\rangle$~\cite{Amin:2022pzv,Jain:2023ojg,Chen:2023bqy}. Thus our lower limit applies equally to $s=0$ and to higher spin DM. 

There are a number of studies of the Milky Way satellites and other galaxies, which attempt to fit a model inspired by the behaviour of ultralight bosons, namely an outer power-law piece, with an inner solitonic core~\cite{Schive2014}. The amplitude of the ground state wavefunction in our analysis effectively accounts for the presence of a soliton with an arbitrary ``core-halo'' relation~\cite{Schive2014b,Chan:2021bja} (displacement of the soliton from the halo centre in simulations is small~\cite{Schive2020} and should not affect our spherical halo assumption). For example, Refs.~\cite{Schive2014,Marsh2015,Calabrese2016,Chen2017,Gonzalez2017, Safarzadeh2019, Goldstein:2022pxu, Hayashi2023} also use Jeans analysis or simplifications thereof, while Ref.~\cite{Bar:2018acw} fits rotation curves. All of these works find \emph{preferred} values of $m$ that fit the cored density profiles of certain galaxies. When taken globally, the smallest cores lead to results that are in tension with the observed masses of larger galaxies, leading to lower limits or very large preferred values for $m$ as in Refs.~\cite{Safarzadeh2019, Goldstein:2022pxu, Hayashi2023}. It is well known, however, that cores can be explained by other physics, such as stellar heating~\cite{Pontzen:2014lma}. For the first time in this context, we used a fully self-consistent density profile comprised of energy eigenstates instead of the heuristic composite, and thus our limit is unaffected by uncertainties on the ``core-halo'' relation~\cite{Chan:2021bja}. In essence, our robust limit comes down to fitting the inner slope of a non-cored object, which is therefore also unaffected by any baryonic physics that may lead to cores in other galaxies. 

\vspace{8pt}
\noindent \textbf{Conclusions.} We have developed a robust and advanced methodology, both computationally and statistically, to set the strongest \emph{fundamental} lower limit to date on the DM particle mass, $m>2.2\times 10^{-21}\text{ eV}$. Our methodology could be extended in future to constrain the fraction of DM with $m<2.2\times 10^{-21}\text{ eV}$ by including an additional cold component in the gravitational potential. Such constraints on mixed DM may be competitive with cosmological bounds, and if accurate to $\mathcal{O}(10\%)$ may probe interesting high energy physics models~\cite{Arvanitaki:2009fg,Marsh:2015xka}. This would also relax the only assumption about the composition of DM in our analysis. 

\vspace*{-10pt}
\section*{Acknowledgements}
\noindent 
We acknowledge useful conversations with Valerie Domcke, Miguel Escudero, Sebastian Hoof, Dima Levkov, Jens Niemeyer, and Hans Winther.
The authors would like to thank Charis Pooni for collaboration at an early stage of the development of \textsc{jaxsp}. TZ acknowledges funding from the European Union's Horizon 2020 research and innovation 
programme under the Marie Sk\l{}odowska-Curie grant agreement No. 945371
and support through a Kristine Bonnevie travel grant (University of Oslo).
JA acknowledges funding from the European Research Council (ERC) under the European 
Union’s Horizon 2020 research and innovation programme (Grant agreement No. 864035) and a fellowship from the Kavli Foundation (from October 2024).
DJEM is supported by an Ernest Rutherford Fellowship from the STFC, Grant No.~ST/T004037/1 and by a Leverhulme Trust Research Project (RPG-2022-145). MF was supported by the United Kingdom STFC Grants ST/T000759/1 and ST/T00679X/1. JIR would like to thank the STFC for support from grants ST/Y002865/1 and ST/Y002857/1.

\bibliography{biblio}
\clearpage
\newpage

\title{Dwarf galaxies imply dark matter is heavier than  $\mathbf{2.2 \times 10^{-21}} \, \mathbf{eV}$ \\ \textit{Supplemental Material}}
\maketitle

\twocolumngrid

\setcounter{equation}{0}
\setcounter{figure}{0}
\setcounter{table}{0}
\setcounter{page}{1}
\renewcommand{\theequation}{S\arabic{equation}}
\renewcommand{\thefigure}{S\arabic{figure}}
\renewcommand{\bibnumfmt}[1]{[#1]}
\renewcommand{\citenumfont}[1]{#1}

\section{\label{sec:render} Wave Function Density at Excluded Boson Mass} 
\noindent Although not relevant for the analysis of the main text, we emphasize that a change in boson mass does not just affect the spherically symmetric part, $\langle|\psi|^2\rangle$, of the wave function density, $|\psi|^2=\langle|\psi|^2\rangle + \chi$, but also the interference term $\chi$. In fact, one expects the size of the interference granules encapsulated in $\chi$ to be of size $\lambda_{\text{dB}}\sim m^{-1}$. 

\begin{figure}[htpb]
    \centering
    \includegraphics[width=0.99\columnwidth]{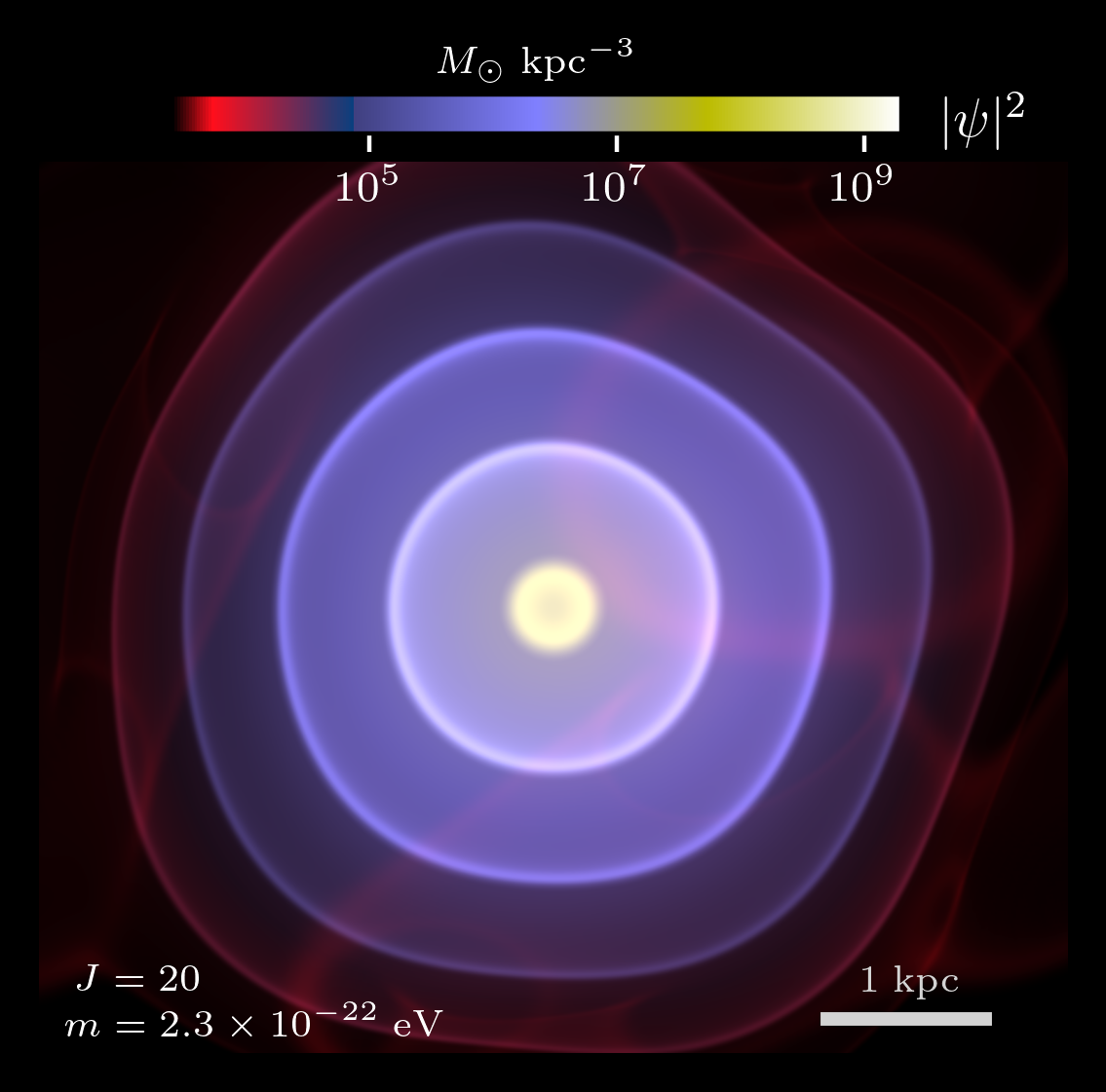}
    \caption{Full wave function as in Fig. \ref{fig:volume_rendering} but for the ten times smaller, excluded mass $m=\SI{2.3e-22}{\electronvolt}$. We find significantly reduced interference due to the $100$ times smaller eigenstate library size of $J=20$.
    }
    \label{fig:volume_rendering_appendix}
\end{figure}

It is instructive to supplement Fig. \ref{fig:volume_rendering} with a wave function reconstruction at a significantly lower mass and assess this behavior qualitatively.
To this end, Fig. \ref{fig:volume_rendering_appendix} depicts the full wave function density $|\psi|^2$ for the same input density $\langle\rho_\text{cNWFt}\rangle$, but at a ten times smaller mass of $m=\SI{2.3e-22}{\electronvolt}$. The result is a density with far less pronounced interference compared to Fig. \ref{fig:volume_rendering}. Note that the number of eigen states entering the mode expansion decreased quadratically to $J = 20$. Fig. \ref{fig:jwkb} shows that for fixed angular momentum $J_l \sim m$ and $J_n \sim m$ at fixed excitation index $n$, recovering $J\sim m^2$. Thus, one naively expects a ten times less modulated wave function in both radial and azimuthal/polar direction respectively --- in accordance with the expectation of the isotropic granule size to follow $\lambda_{\text{dB}}\sim m^{-1}$.

\section{\label{sec:liouville} Bounds from Liouville's Theorem?} 
\noindent The dissipationless dynamics of collisionless particles obey Liouville's
theorem, which states that the phase space distribution, $f(\bm x, \bm p, t)$,
that describes an ensemble of particles, 
remains constant along Hamiltonian trajectories \citep{Binney1987}:
\begin{equation}
    \label{eq:liouville}
    \frac{\text{d}f}{\text{d}t} 
    \equiv \partial_t f + \frac{\bm p}{m} \bm \nabla_x f - m \bm \nabla_x V \cdot \bm\nabla_p f
    =0 \;.
\end{equation}
This is a direct consequence of continuity and the Hamiltonian
equations of motion. An important corollary of Eq. \eqref{eq:liouville} is the
non-decreasing nature of the maximum value of $f(\bm x, \bm p, t)$, $f_\text{max}(t)$: No matter how
complicated the Hamiltonian dynamics between initial and final state may be
(e.g. nonlinear structure formation), the value of $f_\text{max}(t_\text{final})$ cannot exceed its
primordial counterpart $f_\text{max}(t_\text{0})$. Ref.~\cite{Alvey:2020xsk} used this fact to derive stronger,
but model-dependent mass bounds for particle-like fermion dark matter by
reconstructing the value of $f_\text{max}(t)$ in dSphs and comparing it against
DM production mechanism dependent values of $f_\text{max}(t_\text{0})$.

It is then natural to ask if the same strategy may be used to derive a model-dependent
mass limit if DM is assumed to be an ultralight boson. 
Ultralight bosons obey wave-like dynamics, cf. Eq.~\eqref{eq:schroedinger}. 
Consequently, to apply the above argument, we need a (non-relativistic) phase space description of an
ensemble of waves, $f_W(\bm x, \bm p, t)$. Imposing physically natural
conditions on $f_W$,
such as reality, Galilean invariance etc. (see \citep{Hillery1984} for details), Wigner's
distribution,
\begin{equation}
    \label{eq:wigner}
f_W(\bm x, \bm p) = \int \frac{d^3 \bm{x'}}{(\pi \hbar)^3} \exp \left[
\frac{2i}{\hbar} \bm p \cdot \bm{x'} \right] \psi(\bm x - \bm{x'}) \psi^*(\bm x + \bm{x'}),
\end{equation}
is the simplest candidate that satisfies them all \citep{Yourgrau1972}. Taking the time derivative of Eq.
\eqref{eq:wigner} and using Schr\"odinger's equation, we find \citep{Hillery1984}:
\begin{equation}
    \partial_t f_W + \frac{\bm p}{m} \bm \nabla_x f_W= m\sum_{|\bm k| \text{ odd}}
    \frac{\left(\hbar/2i\right)^{|\bm k|-1}}{\bm k!}\frac{\partial^{|\bm
    k|}V}{\partial \bm x^{\bm k}}
    \frac{\partial^{|\bm k|}f_W}{\partial \bm p^{\bm k}}
\end{equation}
with $\bm k \in \mathbb{N_0}$ and $|\bm k| = \sum_i k_i$. 

From the above expression, we see that only a free wave, $V=0$, and a wave trapped in an
harmonic potential can recover $\text{d}f_W/\text{d}t = 0$. A direct
application of the textbook Liouville theorem to the wave scenario for more
general potentials, like in this work, is therefore not possible.

Another common choice is Husimi's distribution $f_H$, a coarse grained, manifestly
non-negative version of Eq. \eqref{eq:wigner}. One may show,
see~\citep{Uhlemann2014}, that $f_H$'s evolution does not obey Liouville's theorem either.

The authors of \cite{Song2024} argue that the failure of the above distribution functions to establish
a wave-analogue of Liouville's theorem lies in the adherence to the symplectic
measure $\text{d}x\wedge\text{d}p$ on phase space. Employing a Haar measure, by contrast, an
incompressible phase-space description can be established, independent of the
specifics of the potential. The extension to systems with infinite degrees of freedom remains open, however.

\section{\label{sec:numerics} Numerical Details of the Analysis Pipeline}
\noindent In the following, we collect definitions, elaborate on numerical techniques and
showcase components of the \textsc{jaxsp} library. 

\subsection{\label{sec:density} Input Density --- \textsc{coreNFWtides}}
\noindent Recall that \textsc{gravsphere} samples the posterior
distribution of the seven parameters that define the flexible \textsc{coreNFWtides}
mass model, $M_{\text{cNFWt}}(r)$ --- an extension of the canonical NFW profile
by a core profile and tidally stripped outskirt. 

The enclosed mass of \textsc{coreNFWtides} profile is given by:
\begin{equation}
    \label{eq:McNFWt}
M_{\text{cNFWt}}(r) =
\begin{cases}
M_{\text{cNFW}}(r) & r < r_t \\
M_{\text{cNFW}}(r_t) & \\
\quad + 4 \pi \rho_{\text{cNFW}}(r_t) r_t^{3 - \delta} & r > r_t\\ 
\qquad \times \left[ \left( \frac{r}{r_t} \right)^{3 - \delta} - 1 \right] & 
\end{cases} \;,
\end{equation}
where the cored mass profile is:
\begin{equation}
M_{\text{cNFW}}(r) = M_{\text{NFW}}(r) f^n \;,
\end{equation}
with:
\begin{equation}
M_{\text{NFW}}(r) = M_{200} g_c \left[ \ln \left( 1 + \frac{r}{r_s} \right) -
\frac{r}{r_s} \left( 1 + \frac{r}{r_s} \right)^{-1} \right] \;,
\end{equation}
and:
\begin{equation}
f^n = \left[ \tanh \left( \frac{r}{r_c} \right) \right]^n\;.
\end{equation}
$\rho_{\text{cNFW}}$ is then given by:
\begin{equation}
\rho_{\text{cNFW}}(r) = f^n \rho_{\text{NFW}} + \frac{n f^{n-1} (1 - f^2)}{4 \pi
r^2 r_c} M_{\text{NFW}} \;,
\end{equation}
which modifies the standard NFW density:
\begin{equation}
\rho_{\text{NFW}}(r) = \rho_0 \left( \frac{r}{r_s} \right)^{-1} \left( 1 +
\frac{r}{r_s} \right)^{-2} \;.
\end{equation}
The central density $\rho_0$ and scale radius $r_s$ are given by:

\begin{equation}
\rho_0 = \rho_{\text{crit}} \Delta c_{200}^3 g_c / 3\;, \quad r_s =
\frac{r_{200}}{c_{200}} \;,
\end{equation}
where we defined:
\begin{equation}
g_c = \frac{1}{\log (1 + c_{200}) - \frac{c_{200}}{1 + c_{200}}} \;,
\end{equation}
and set the virial radius to be:
\begin{equation}
r_{200} = \left[ \frac{3}{4} M_{200} \frac{1}{\pi \Delta \rho_{\text{crit}}}
\right]^{1/3} \;. 
\end{equation}
$c_{200}$ denotes the dimensionless concentration parameter, $\Delta = 200$ and
$\rho_{\text{crit}} = 136.05 M_\odot \text{kpc}^{-3}$ is the present day
critical density of the Universe. $M_{200}$ is the mass within the virial radius $r_{200}$. 

\begin{figure}[htpb]
    \centering
    \includegraphics[width=0.99\columnwidth]{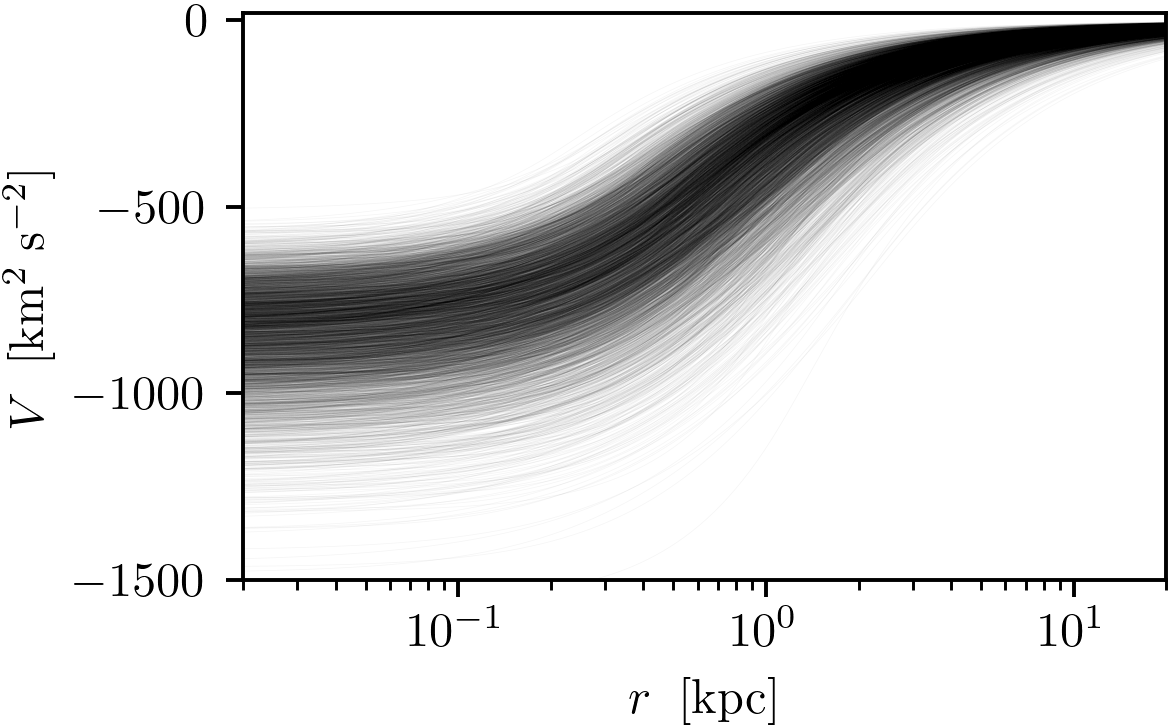}
    \caption{Ensemble of gravitational potentials of the \textsc{gravsphere} posterior from
    Fig. \ref{fig:gravsphere_posterior} obtained via Eq. \eqref{eq:potential}.}
    \label{fig:potential}
\end{figure}

With an analytic expression for the enclosed mass up to radius $r$ available, we
can forgo solving Poisson's equation and obtain the gravitational potential,
\begin{equation}
    \label{eq:potential}
    V(r)  = -G \int_r^\infty \text{d}s \frac{M_\text{cNFWt}(s)}{s^2} \;,
\end{equation}
via high order Gauss-Legendre quadrature.
Fig. \ref{fig:potential} depicts the ensemble of potentials implied by the posterior distribution
shown in Fig. \ref{fig:gravsphere_posterior}.

Our analysis relies exclusively on \textsc{coreNFWtides}, and is thus the only
mass model currently implemented in \textsc{jaxsp}. Introducing alternative
choices, however, is trivial.

\subsection{\label{sec:fit} Construction of Eigenstate Libraries}
\noindent The task of constructing the eigenstate libraries amounts to diagonalizing:
\begin{equation}
    \begin{split}
    \label{eq:schroedinger_appendix}
    -\frac{\hbar^2}{2m}\left(\frac{\partial^2}{\partial r^2}
    -\frac{l(l+1)}{r^2}\right)u_{nl} + m V u_{nl}&=E_{nl} u_{nl} \;, \\
    u_{nl}(0) = u_{nl}(\infty) &= 0
    \end{split}
\end{equation}
for eigenmodes $\psi_{nlm}(\bm x,t) = r^{-1}u_{nl}(r)Y^m_l(\phi,\theta)$.
Once a mode $u_{nl}$ penetrates the bounding gravitational potential at a radius $\tilde r$,
it undergoes exponential decay into the classically forbidden region:
\begin{equation}
    u_{nl}(r) \sim \exp \left( -\frac{m}{\hbar}
    \int_{\tilde r}^{r} \mathrm{d}s \sqrt{2(E_{nl} -V_{\text{eff}}(s)) }\right)\;.
\end{equation}
To accommodate all modes with energy less than $E$ on the same grid, Ref.~\cite{Ledoux2006} suggests setting the 
upper domain limit $r_\text{max}$ according to:
\begin{equation}
    \label{eq:truncation}
    \frac{m}{\hbar} \int_{\tilde r}^{r_{\text{max}}} \mathrm{d}s
    \sqrt{2(V_{\text{eff}}(s) - E)} = 18\quad\text{with}\quad V(\tilde r)
    = E \;,
\end{equation}
such that $\partial_r u_{nl}(r_\text{max}) = 10^{-16}u_{nl}(r_\text{max})$ and
thus negligible at the level of floating point precision. We adopt this physically inspired domain truncation for our considerations.

The Leo II eigenstates implied by Eq. \eqref{eq:schroedinger_appendix} usually
spread over 2-3 orders of magnitude in spatial extent and include power law like
behaviour, $u_{nl} \sim r^{l+1}$,
at small radii, exponential decay in the classically forbidden region as
$r \to \infty$ and, depending on the quantum number $n$, a highly oscillatory,
intermediate regime. To capture these regimes equally well, it has proven
useful to transform to a log linear grid, $x = a r + b \log r$, thus
converting Eq.~\eqref{eq:schroedinger_appendix} into a \emph{non-singular} Sturm-Liouville
problem of the form \cite{Bowen2023}:
\begin{equation}
   \label{eq:sl} 
   \begin{split}
       E_{nl}wt_{nl} &= \left(-\partial^2_x + q_{l}\right) t_{nl}\;,\\ 
       q_{l} &= \frac{\hbar^2}{m}\frac{2}{(b+ar)^2}\left[\frac{l(l+1)}{2} +
       \frac{b(b+4ar)}{8(b+ar)^2}+ \frac{m^2}{\hbar^2}r^2V \right] \;, \\
       w &=  \frac{2r^2}{(b + ar)^2}\;, \\
       t_{nl}(x) &= u_{nl}\left(r(x)\right)\sqrt{\frac{b}{r} + a}\;, \\
   \end{split} 
\end{equation}
where $r(x) =\frac{b}{a}W\left(\frac{a}{b}e^\frac{x}{b}\right)$ and $W(x)$
is the Lambert W function. In practice, we set $a=1$ and $b=10$ for all
computations in the main text, although more refined strategies are conceivable.

A plethora of methods exist to solve Eq. \eqref{eq:schroedinger_appendix} or
\eqref{eq:sl} respectively. In the context of
reconstructing wave functions for ULDM haloes with $m\lesssim
\SI{1e-22}{\electronvolt}$ \citep{Yavetz2022} or low total mass systems like ultra faint dwarfs
\cite{Dalal2022}, finite
difference discretisations and an eigen-decomposition of the Hamiltonian matrix $\bm H_l$
have proven to be convenient. Indeed, matrix methods are appealing due to their
exceptional ease of implementation. 

An intrinsic limitation of this textbook approach is that the
discretization must satisfy:
\begin{equation}
    \label{eq:rank}
    \mathrm{rank}(\bm{H}_l) > c \cdot J_l(E), \quad c=\mathcal{O}(1-10).
\end{equation}
where $J_l(E)$ denotes the number of eigenstates at angular momentum
$l$ up to energy $E$ and $c$ is a discretisation dependent constant that assures sufficient resolution of 
each mode, especially for the highly excited, oscillatory states. In practice, we
find $c\gtrapprox 8$ to be required to suppress spurious, unphysical modes for
centered finite differences (\texttt{centered\_fd}) and $c\gtrapprox3$ for a Chebyshev collocation approach.
The case $c=1$ represents the fundamental requirement on $\bm{H}_l$'s orthonormal eigenbasis
to span a space of at least $J_l$ dimensions in order to recover $J_l$ modes. 
Note that the diagonalisation of $\bm{H}_l \in \mathbb{R}^{N \times N}$
takes $\mathcal{O}(N^3)$ time and while higher order discretisations reduce
the value of $c$, they also lead to less sparse, non-hermitian matrices demanding
more general and thus less performant diagonalisation techniques.

We can estimate $J_l(E)$ as the number of roots of the phase function in a WKB ansatz for $u_{nl}$:
\begin{equation}
    \label{eq:wkb}
    J_l(E) \simeq \left\lceil \frac{1}{\pi}
        \int_0^r\text{d}s\sqrt{\frac{2m^2}{\hbar^2}\left(E - V(s)
        -\frac{l(l+1)}{s^2}\right)}\;\right\rceil\;,
\end{equation}
with $\lceil.\rceil$ rounding to the next biggest integer.
To assess whether Eq. \eqref{eq:rank} would pose a fundamental obstacle for the
analysis in the main text, we pick the deepest potential from Fig.
\ref{fig:potential}, which translates to picking the largest number of eigenstates, and
substitute it alongside $E=V\left(r_{99}\right)$ and $r=r_{99}$ into Eq.
\eqref{eq:wkb}. We then set $c=10$, $l=0$, and plot the right hand side of Eq.
\eqref{eq:rank} to estimate a conservative, lower bound on the matrix rank. 

\begin{figure}[t]
    \centering
    \includegraphics[width=0.99\columnwidth]{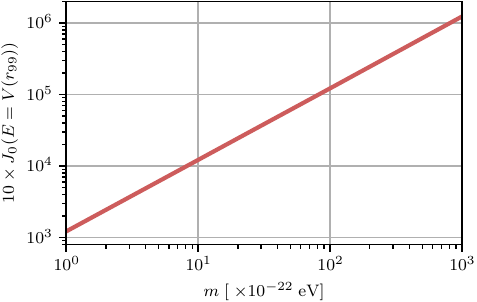}
    \caption{WKB-based estimate for the minimal rank of a finite difference discretised
    Hamiltonian $\bm H_l$ for the deepest potential well in Fig.
\ref{fig:potential} as function of boson mass $m$.}
    \label{fig:jwkb}
\end{figure}

Fig. \ref{fig:jwkb} depicts the result. We find $\mathrm{rank}(\bm H_0) >
3\times10^{4}$ at $m=\SI{2.3e-21}{\electronvolt}$ (the first accepted mass
according to our analysis). Note that for a fully exhaustive, 
eigenstate library construction as in the main text, the angular quantum number $l$
needs to be increased to $l=L$ at which $J_L(E) = 0$. In other words, $L$
diagonalisations need to be carried out in total. 
For the above parameter choice $L\simeq 3\times10^4$. Since each
diagonalisation is independent of all the other, they may, given sufficient
compute resources be carried out in parallel. We thus concur that, while
challenging, a second order, centered difference approach (yielding a
tridiagonal $\bm H_l$), may still be a viable option, if condition
in Eq.~\eqref{eq:rank} is the only concern. 

Another drawback, also reported in \cite{Pryce1993}, is the singular nature of the
angular momentum barrier in Eq. \eqref{eq:schroedinger_appendix} and stiffness of
Eq. \eqref{eq:sl}, translating to high condition numbers of $\bm{H}_l$ and consequently
degraded convergence. We have found that the severity of this issue is
problem, i.e. potential, dependent.
Recall that our analysis relies on the library construction for thousands of different
potentials making individual convergence checks tedious to carry out. Thus, a
scheme less prone to the stiffness of Eqs. \eqref{eq:schroedinger_appendix} - \eqref{eq:sl}, 
would be more suited for this automated, ``problem agnostic'' approach.

Moreover, note that for matrix methods, all modes of fixed angular momentum must live on the 
same spatial grid irrespective of the discretisation used and thus depriving us of the
possibility to optimise the mode quality based on their actual spatial extent.

Additional short-comings, most notably poor error bounds for excited eigenvalues and modes, are summarised in \cite{Pryce1993} and lead to the conclusion
that applying matrix methods in the context of
automated diagonalisation of singular problems, and up to high energies, to be challenging.

With our robust bulk diagonalisation task in mind, but also in preparation for applying wave function 
reconstructions to (i) higher mass systems
(both in terms of total and boson mass) and (ii) non spherical objects, 
we have implemented a piecewise perturbative approach in \textsc{jaxsp}. This
approach, also known as Pruess' method \cite{Pryce1993}, may be seen as a simple
representative of the more powerful class of constant perturbation methods,
with \citep{Ledoux2010} building the backbone of general purpose
Sturm-Liouville codes such as \cite{Ledoux2016}.

The key idea is simple:
Instead of numerically approximating the solution of the exact Sturm-Liouville problem \eqref{eq:sl}, 
we introduce an approximation at the equation level and solve the surrogate equation to 
desired accuracy. More precisely, we partition the whole domain $\Omega$ into $N$ equally
sized sectors $\Omega = \cup_{k=1}^N \Omega_k = \left[x_{k-1}, x_{k-1} + \Delta x\right]$ and approximate 
the coefficient functions $w(x)$ and $q_l(x)$ of Eq. \eqref{eq:sl} as piecewise
constant functions, i.e. $Q_l(x \in \Omega_k) = q_l\left(\frac{1}{2}(x_{k-1} + x_{k})\right)$. 
The solution, $T_k$, for this Sturm-Liouville with piecewise constant coefficient functions
on $\Omega_k$ takes the form:
\begin{widetext}
\begin{equation}
    \label{eq:reference}
    T_k(x, E) = 
    \begin{cases}
        T_k(x_{k-1}, E) \frac{\sin\left(\omega_k (x_k - x)\right)}{\sin\left( \omega_k \Delta x\right)}
        + 
        T_k(x_{k},E) \frac{\sin\left(\omega_n (x - x_{k-1})\right)}{\sin\left( \omega_k \Delta x\right)}
        &\omega^2_k \equiv W_k E - Q_k > 0 \\
        T_k(x_{k-1}, E) \frac{\sinh\left(\sqrt{|\omega^2_k|} (x_k -
        x)\right)}{\sinh\left(\sqrt{|\omega_k|^2} \Delta x\right)}
        +
        T_k(x_{k}, E) \frac{\sinh\left(\sqrt{|\omega^2_k|} (x
        -x_{k-1})\right)}{\sinh\left(\sqrt{|\omega_k|^2} \Delta x\right)}
        &\omega^2_k < 0
    \end{cases} \;.
\end{equation}
\end{widetext}

An approximation to the original problem in Eq.~\eqref{eq:sl} may then be found by
stitching together all sectors and demanding the full solution $T \in
C^1(\Omega)$. Compatibility of the sector solutions at the interface points
$\bm T~=~\begin{pmatrix} T_0(x_0, E), \dots, T_{n-1}(x_{n-1},E) \end{pmatrix}^\intercal $ then imply
\cite{Milkhailov1983}:
\begin{equation}
    \label{eq:transcendental}
    \bm K(E) \bm T = 0 \;,
\end{equation}
with a tridiagonal matrix $K$ of non-polynomial element functions of the yet to
be determined eigen energies $E$. A brute force way of obtaining these is to demand:
\begin{equation}
    \label{eq:brutforce}
    \mathrm{det}(\bm K(E)) = 0 \;,
\end{equation}
and solve for all roots smaller than the desired cutoff energy. This should be compared
with the matrix method version of Eq. \eqref{eq:brutforce},
\begin{equation}
    \label{eq:matrix}
    \mathrm{det}(\bm{H}_l - E\mathbb{1}) = 0 \;.
\end{equation}
Eq. \eqref{eq:matrix} is a polynomial equation of degree $N$, and thus contains $N$
roots. Finding more roots necessitates an increase in $N$. The transcendental
nature of Eq. \eqref{eq:brutforce}, by contrast, contains an infinite number of
roots, independent of $N$ and will therefore always yield an approximation to
any desired state of Eq. \eqref{eq:sl} 
\footnote{
    The mode and eigenvalue quality obviously depends on the number of sectors N.
}.

Obtaining the eigenvalues via Eq. \eqref{eq:brutforce} is numerically
ill-conditioned: $\mathrm{det}(\bm K)$ is characterised by rapid oscillations and
enormous gradients which make root finding challenging. Instead, we employ the Wittrick-Williams algorithm
\citep{Wittrick1971, Milkhailov1983, Yuan2004} stating that the \emph{exact} number of
eigen modes $J(E)$ up to an arbitrary energy $E$ may be decomposed into two
contributions:
\begin{equation}
    \label{eq:ww}
    J(E) = J_0(E) + s(\bm K(E)) = \sum_{k=1}^N \left\lfloor\frac{\omega_k\Delta
    x}{\pi}\right\rfloor + s(\bm K(E)) \;,
\end{equation}
where $J_0(E)$ denotes the number of eigen modes up to energy $E$ if each domain
sector obeys Dirichlet conditions individually and $s(K(E))$ is the ``sign count"
of $K(E)$. Note that the second equality only holds in the case of the sector
solution in Eq.~\eqref{eq:reference}. 
$s(\bm K(E))$ may be obtained in $O(N)$ time for tridiagonal matrices
like in the present case. 

Eq. \eqref{eq:ww} shows staircase-like behavior. The value of the
sought-after eigen-energies coincide with the sites where J(E) undergoes a unit jump. Thus
we may bound eigen value $E_n$ in the interval $[E_l, E_u]$ with assured
convergence by simply bisecting $J(E) - n$ on $[0, E]$ and setting $E \simeq
\frac{1}{2}\left(E_u + E_l\right)$.

\begin{figure*}
    \centering
    \includegraphics[width=0.99\textwidth]{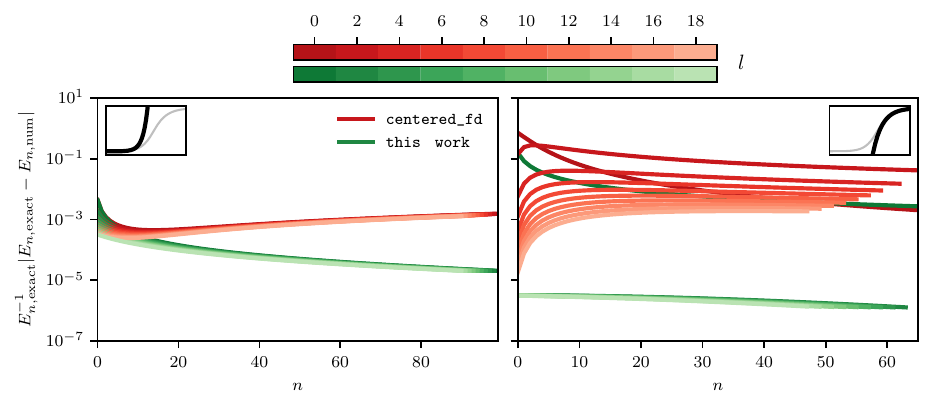}
    \caption{Relative error of the eigen energies $E_{nl}$ as a function of
    excitation index $n$ and angular momentum $l$ (color shade) in case of an
    harmonic (\textbf{left}) and Coulomb potential (\textbf{right}). The
    second order, cf. Fig. \ref{fig:order} piecewise perturbative appproach of this work (green) 
    consistently outperforms its centered difference, matrix method competitor
    in both cases.
    }
    \label{fig:eigenvals}
\end{figure*}

We illustrate the above method in Fig.~\ref{fig:eigenvals} in terms of the
eigenvalues for two approximation of the potential of the mean posterior
density at the accepted mass $m=\SI{2.3e-21}{\electronvolt}$
for which analytic spectra are available: At small radii, $V(r)$ remains
finite. Introducing a curvature parameter $\omega^2$ thus allows for an isotropic,
harmonic oscillator approximation:
\begin{equation}
    \label{eq:osci_spectrum}
    V_\text{osci}(r) = \frac{1}{2}\omega^2r^2\;,\quad E_{\text{osci},nl} =
    \frac{2}{3}E_{\text{osci},00}\left(2n + l + \frac{3}{2}\right) \;.
\end{equation}
For large $r$, a monopole approximation is adequate:
\begin{equation}
    \label{eq:Hspectrum}
    V_\text{H}(r) = -\frac{GM_\text{tot}}{r}\;,\quad E_{\text{H},nl} =
    -\frac{E_{\text{H},{00}}}{(n+l+1)^2} \;.
\end{equation}
Fig. \ref{fig:eigenvals} shows the relative eigenvalue error as a function of the energy index $n$ relative
to the exact spectra of Eq. \eqref{eq:osci_spectrum} or \eqref{eq:Hspectrum}
respectively.
In both cases, our approach (green) outperforms the textbook matrix
diagonalisation approach (red). Shades of green/red designate different angular
momenta $l\in [0, 20]$. We see at worst uniform accuracy for our method as index
$n$ increases, and no significant dependence on $l$. The latter
is a consequence of the coordinate transformation leading to Eq.
\eqref{eq:sl}, converting the challenging angular momentum barrier into a
trivial offset. The degraded accuracy for $l=0$ is a result of this
transformation as well. Note that $x=ar+b\log r$ pushes the origin to negative
infinity where Dirichlet conditions are assumed. This is impossible on a finite
grid and for $l=0$, $u_{nl} \sim r$ violates this behavior most dominantly. 
Note that we tried explicitly to solve Eq. \eqref{eq:sl} with
\texttt{centered\_fd} as well, leading to unstable results across different
potentials, which aligns with our previous comments on the behavior of \texttt{centered\_fd}
for singular/high-condition number problems.

With a bounding interval $(E_l, E_u)$ for the eigenvalue $E_n$ at hand, we adopt
the mode-finding approach of \cite{Yuan2004}: Let $E_m = \frac{1}{2}\left(E_u +
E_l\right)$ be the approximation to $E_n$. One may then show that the
eigenpair with the smallest eigenvalue $\mu$ that solves the generalised eigen problem
\begin{equation}
    \bm K(E_m) \bm T = \mu \left(\bm K(E_u) - \bm K(E_l)\right) \bm T
\end{equation}
approximates $\bm T(E_n)$ to high accuracy, i.e. effectively with error $\sim
(E_m - E_n)^2$. We are guaranteed to obtain a well-converged approximation to
this pair via inverse iteration, which due the simple, sparse structure of $\bm
K$ takes $\mathcal{O}(N)$ steps.

At last, the mode $u_{nl}$ may be normalised by computing $N^2 =
\sum_{k=1}^Nw_k\int_{x_{k-1}}^{x_k}\text{d}x T_k^2(x, E_m)$, analytically \cite{Milkhailov1983}. 

\begin{figure}
    \centering
    \includegraphics[width=0.99\columnwidth]{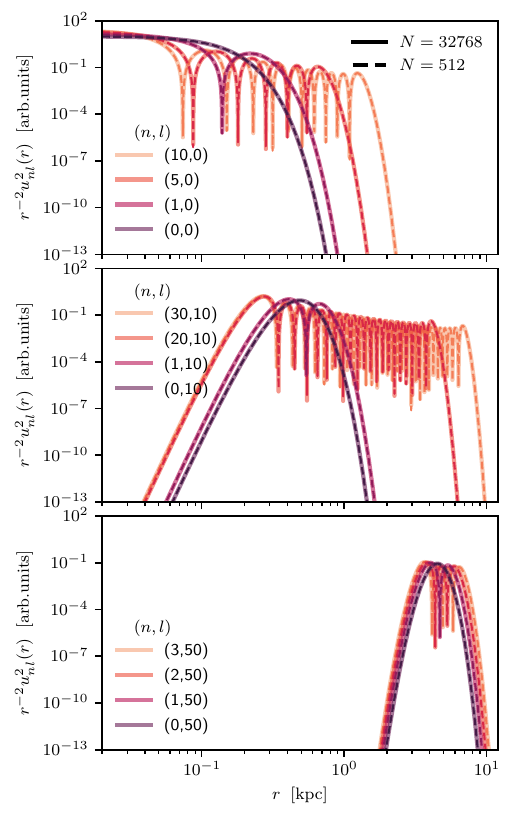}
    \caption{Excerpt of the eigenstate library for the mean density at
        $m=\SI{2.3e-21}{\electronvolt}$. Shown are eigenmodes at low (\textbf{top}), moderate
        (\textbf{middle}) and high (\textbf{bottom}) angular momentum $l$ at two grid
        resolutions. Across all modes, we observe no significant differences between $N=512$ and
        the 64 times higher reference solution with $N=32768$, indicating a
        robust and well converged eigenstate library.
    }
    \label{fig:eigenmodes}
\end{figure}

Fig. \ref{fig:eigenmodes} illustrates an excerpt of the eigenstate library for two
grid resolutions: $N=512$ (dashed) and $N=32768$ (solid). Depicted is a selection of
low/high angular momentum modes for a range of excitation numbers $n$. In all
cases, we find satisfying convergences already at a comparatively low sector
count of $N=512$. In fact, no significant difference to the case of $N=32768$ is
observable. Intuitively, we attribute this behavior to the explicit treatment of the
mode oscillation in Eq. \eqref{eq:reference}, effectively fixing the small scale
features of each mode at the equation level. What remains is to determine the
large scale behavior of each eigenstate --- a task that can then be solved at a
reduced spatial resolution. Put differently, the stiffness of Eq. \eqref{eq:sl} is a
non-issue in our approach.

\begin{figure}
    \centering
    \includegraphics[width=0.99\columnwidth]{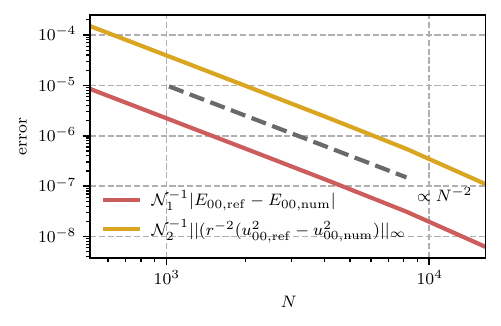}
    \caption{%
        Order of convergence of the $n=l=0$ eigen energy (red) and eigen state
        (yellow) as a function of grid resolution and relative to a reference
        solution with $N=32768$. The normalisation constants are
        $\mathcal{N}_1=E_{00,\text{ref}}$ and $\mathcal{N}_2 = {r^*}^2 u_{00,\text{ref}}^2(r^*)$ 
        with $r^* = \text{argmax}|r^{-2} (u^2_{00,\mathrm{ref}}-u^2_{00,\mathrm{num}})|$.
        Quadratic convergence is recovered.
    }
    \label{fig:order}
\end{figure}

Fig. \ref{fig:order} shows our method's order of convergence for the ground state mode of
the full potential (itself the most dominant contribution in the full wave function expansion) as a function of
sector count $N$ with respect to a reference solution at $N=32768$. We find both
eigenvalue (red) and eigenmode (yellow) to converge quadratically.

The piecewise constant perturbative approach offers additional advantages worth
mentioning:
Firstly, note that the construction of an individual mode happens independent
of all other modes. Thus, this allows for a trivial parallelisation of the
entire eigenstate library construction but also introduces the possibility of
per-mode grid optimisations. At present, we do not leverage this flexibility.
However, a simple extension, which we mention only in passing but intend to 
explore in the future is a two-pass, predictor-corrector-like approach: 
In a first stage we approximate the eigenvalue
$E_n$ by adhering to the global domain truncation of Eq. \eqref{eq:truncation}.
For the second stage, we re-set $\tilde r$ to be radius satisfying $V(\tilde r) = E_m$ (which is
easily found by locating a sign-flip in the list of $\omega_k$'s), and re-compute
$r_\text{max}$ via Eq. \eqref{eq:truncation} with an increased n-folding
number. The result is an improved grid resolution for modes with smaller spatial
extent. 
Note that such per mode optimizations are not possible with matrix
methods. A per-mode construction also paves the way for incorporating prior
knowledge about the significance of eigenmode $(n,l)$ for the final fit: The
authors of \cite{Yavetz2022} show
that the mode coefficient $|a_{nl}|^2$ is asymptotically equivalent to the equilibrium phase space
distribution. Thus, constructing the latter beforehand can be
used to cherry-pick modes based on the value of the phase-space distribution,
thereby reducing the total time required to construct the significant part of
the eigenstate library.

We emphasize we have not exhausted all possible optimisations in the
current iteration of \textsc{jaxsp} as the application of the main text did not
demand for it. We close this section by mentioning two of them. 
Firstly, and as
outlined in \cite{Yuan2003}, substantial runtime savings may be gained by relaxing the
initial eigenvalue bound $(E_l, E_u)$ but sharpening both eigenmode and
value simultaneously --- effectively reducing the number of bisection steps and
evaluations of Eq. \eqref{eq:ww}. 
Secondly, employing a non-uniform domain decomposition,
e.g. \cite{Yuan2014}, may reduce the libraries memory footprint while maintaining accuracy. We
anticipate both aspects to be relevant for an extension of the stationary
ultra-light dark matter approximation to higher mass systems.

\subsection{\label{sec:fit} Fitting the Input Density}
\noindent Recall that self-consistency requires us to find superposition coefficients $a_{nl} \in \mathbb{C}$
such that the input density $\rho_\mathrm{in}$ is recovered:
\begin{equation}
    \label{eq:self-consistency}
    \rho_\mathrm{in}(r) 
    \stackrel{!}{=} 
    \langle |\psi|^2\rangle = (4\pi r^2)^{-1}\sum_{nl} (2l+1)|a_{nl}|^2
    u_{nl}^2(r) \;.
\end{equation}
A variety of quasi-distance measures can in principle be employed to minimise
the discrepancy between the left and right hand side of Eq.
\eqref{eq:self-consistency}. For example, Refs.~\cite{Yavetz2022, Gosenca2023} use a square distance measure:
\begin{equation}
    \label{eq:square_distance}
    d_\mathrm{sq}(\rho_\mathrm{in}, \langle|\psi|^2\rangle)
    = \frac{1}{r_\mathrm{fit}}\int_0^{r_\mathrm{fit}}\text{d}r \left(\frac{\langle|\psi|^2\rangle -
    \rho_\mathrm{in}}{\rho_\mathrm{in}}\right)^2 \;,
\end{equation}
while this work employs the symmetric Jensen-Shannon divergence (JSD):
\begin{equation}
    \begin{split}
        \label{eq:jsd}
    d_\mathrm{JSD}(\rho_\mathrm{in}, \langle|\psi|^2\rangle)
    &= \frac{1}{2} \left[d_\mathrm{KL}(\rho_\mathrm{in}, m) +
    d_\mathrm{KL}(\langle|\psi|^2\rangle, m)\right]\\
    \text{with\qquad}m &= \frac{1}{2}\left(\rho_\mathrm{in}+\langle|\psi|^2\rangle\right)
    \end{split} \;,
\end{equation}
and $d_\mathrm{KL}$ as the Kullback-Leibler divergence:
\begin{equation}
\label{eq:kl}
d_\mathrm{KL}(p, q) = 4\pi\int_0^{r_\mathrm{fit}} \text{d}rr^2 p(r)
\log_2\left(\frac{p(r)}{q(r)}\right)\;.
\end{equation}
Our choice for JSD is motivated by its convenient theoretical properties,
namely symmetry, $d_\mathrm{JSD}(p,q) = d_\mathrm{JSD}(q,p)$, and boundedness, 
$0 \leq d_\mathrm{JSD}(p,q) \leq 1$. 

While Eq. \eqref{eq:square_distance} is
neither symmetric nor bounded from above, a property common with
Eq. \eqref{eq:jsd} is that positivity, $d>0$, can be
established even if $p$ and $q$ have differing total masses, $M_p \neq M_q$. This
is trivial to see for Eq. \eqref{eq:square_distance}. To see this in the JSD case, we use the fact that $\log(x)
\leq x - 1$ to get $d_\mathrm{KL}(p, q) \geq M_p - M_q$ and thus:
\begin{equation}
    \begin{split}
    2 d_\mathrm{JSD}(p,q) 
    &\geq M_p - M_{\frac{p+q}{2}} + M_q - M_{\frac{p+q}{2}} \\
    &\geq M_{p+q} - M_{p+q} = 0 \;. 
    \end{split}
\end{equation}
This is a clear advantage over using KL as an objective function, for which an unconstrained set of
trial coefficients $a_{nl}$ may lead to $d_\mathrm{KL} < 0$ during the optimisation and thus divergence from the
solution $p=q$ where $d_\mathrm{KL}(p,q) = 0$. Put differently, when using
either Eq. \eqref{eq:square_distance} or \eqref{eq:jsd}, every point in
the space of coefficients $\bm a_{nl} \in \mathbb{C}^J$ is associated with a
positive distance value, which allows a descent to the global minimum even if no
complex simplex constraint of the form $\bm a_{nl}^\intercal \bm a_{nl} = 1\;,
a_{nl} \ge
0\;\forall (n,l)$ is implemented.

Starting from randomised intitial conditions, we employ \texttt{jaxopt}'s
gradient descent as a warm up minimiser, followed by LBFGS until 
$\epsilon = ||\nabla_{\log\left(|a_{nl}|^2\right)} d|| < 10^{-7}$.
We optimise $\log\left(|a_{nl}|^2\right)$ which effectively
standardises the scale of all involved features, improves convergence and
simultaneously implements the positivity constraint, $|a_{nl}|^2>0$, on all mode coefficients.

Clearly, the specifics of the optimal coefficient set will depend on the choice of
objective function. There is no reason to believe $|a_{nl,\mathrm{JSD}}|$ is
identical to $|a_{nl, \mathrm{sq}}|$ for all modes $(n,l)$, especially in the halo outskirts, where the presence
of a high number of modes may induce degeneracies. At low radii, however, fewer
modes exist, and their respective weights are higher. It seems reasonable to
expect that these dominant coefficients are roughly similar across different objective
functions. 

\begin{figure}[htpb]
    \centering
    \includegraphics[width=.99\columnwidth]{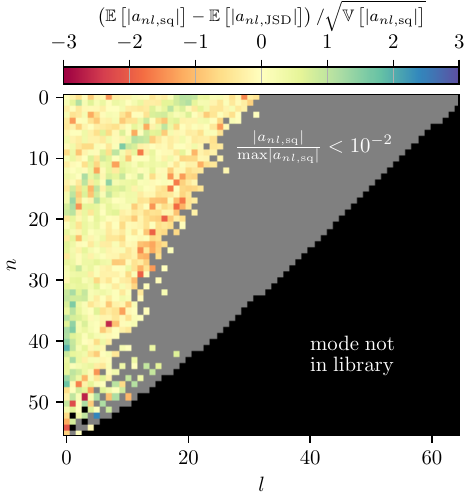}
    \caption{Per-mode comparison between two sets of $16$ wave functions
        obtained with either Eq. \eqref{eq:square_distance} or Eq.
        \eqref{eq:jsd} and random initial conditions.
        We find the mode coefficients of both sets to be 
        consistent within $3\sigma$ of the square distance set standard error.}
    \label{fig:objective}
\end{figure}
We illustrate this in Fig. \ref{fig:objective} comparing the mode coefficients $a_{nl}$ obtained from
minimising Eq. \eqref{eq:square_distance} or \eqref{eq:jsd} starting from a
sample of random initial conditions.
More precisely, we take the JSD minimising wave function
coefficients of Fig. \ref{fig:volume_rendering} as ground truth
input density and reconstruct a set of $16$ wave functions with either Eq.
\eqref{eq:square_distance} or \eqref{eq:jsd} and different random initial
conditions. We expect the most dominant modes, i.e. modes with a weight
of at least $1\%$ of $\mathbb{E}[|a_{nl, \mathrm{JSD}}|]$, to be comparable on
average. Fig. \ref{fig:objective} supports this assertion by measuring the deviation of the per-mode 
sample means relative to its standard error. We find all coefficients to be within $3\sigma$ of
$\mathbb{E}[|a_{nl, \mathrm{JSD}}|]$.

\begin{figure}[htpb]
    \centering
    \includegraphics[width=.99\columnwidth]{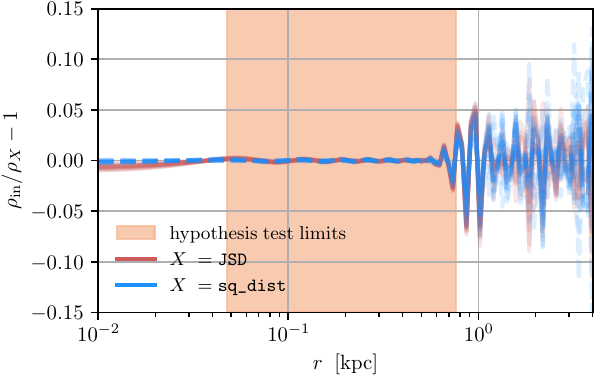}
    \caption{Relative deviation of the set of $16$ JSD (red) and square distance
    fits (blue) from the input density. In all cases the error is at the
    sub-percent level within the limits of the hypothesis test, supporting
    the robustness of our analysis with respect to details of the coefficient optimisation.}
    \label{fig:objective_densities}
\end{figure}

In practice, only $\langle|\psi|^2\rangle$ matters for our analysis. Fig.
\ref{fig:objective_densities} therefore compares both sets of wave function
densities directly. We find only sub-percent level deviations within the limits of
the hypothesis test, suggesting that the optimsiation and its details do not induce a
systematic source of error in our analysis.

\subsection{\label{sub:statistics} Statistics}
\noindent Acknowledging our deviation from a textbook bayesian Jeans analysis, we provide 
a pedagogical introduction to our statistical framework, with the intent of building 
additional intuition for the robustness of our limit.
We then summarize technical elements of the hypothesis test.
For a more detailed discussion of the latter, we refer to \cite{Gretton2012,Wynne2020,Biggs2023}.

\subsubsection{\label{subsub:estimate}Intuition from sample-based estimates}
\noindent Setting aside technical intricacies, what mass limit can we expect from a conservative statistical approach? 
Insight on this can be gained by observing the wave function fit quality for extreme 
draws within the posterior samples.
For this we sort the set of \textsc{gravsphere} draws according to their eigen state library size $J$ and consider two limiting cases: (A) A strongly cusped profile producing a minimal
value of $J$ and (B) a highly cored density associated with a maximal value of $J$. The additional condition on the library size is important as a reduced/increased number of modes impedes/improves the model flexibility at fixed boson mass. Thus, we expect case A, i.e. a highly inflexible wave function, fitting a challenging density, to provide a stronger limit compared to case B, for which a large number of modes is already available at low mass to reproduce a simple (cored) envelope density.

\begin{figure}[htpb]
    \centering
    \includegraphics[width=.99\columnwidth]{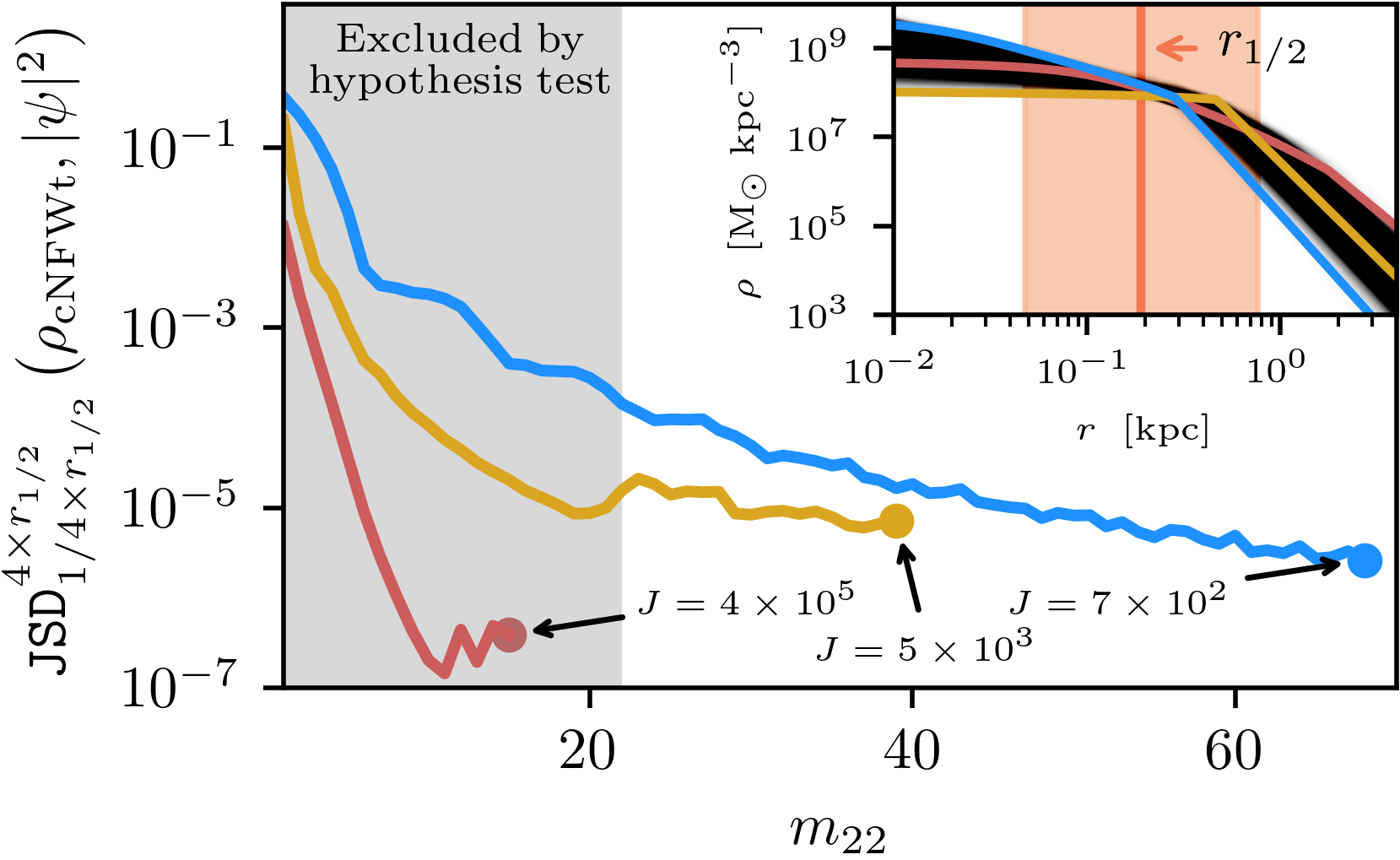}
    \caption{%
        Value of the JSD for extreme densities within the \textsc{gravsphere} posterior (\textbf{inset}) as a function of boson mass. We distinguish between a highly cored density given a maximal eigen state library size (red), a strongly cusped realization for a minimal eigen state library size (blue) and the draw with the most extended core (yellow).
    }
    \label{fig:simple_estimate}
\end{figure}
Fig. \ref{fig:simple_estimate} depicts the behavior of the JSD for both cases 
($A=\text{blue}$, $B=\text{red}$) as a function of boson mass. 
In general, we find the value of the objective function to quickly decay over multiple orders of magnitude until a convergence plateau at $m_{22,A} \gtrsim 60$ and $m_{22,B}\gtrsim10$ is attained. The chosen draws constitute extreme cases of the ensemble, such that any other random sample considered in isolation will realize $m_{22,B} \leq m_{22} \leq m_{22,A}$. We exemplify this with the most cored realization (unconditioned on $J$) in yellow, which converges close to our mass limit in the main text.

\subsubsection{From a sample estimate to an ensemble estimate}
\noindent It is clear from the discussion around Fig. \ref{fig:simple_estimate} that a mass of
$m_{22}\gtrsim 60$ is able to reproduce the \emph{entire} \textsc{gravsphere} 
posterior, while a mass below the
conservative estimate of $m_{22}\gtrsim 10$ would reproduce \emph{none} of the posterior
samples. Our result presents a robust compromise between both extremes by
testing for a weaker notion of convergence. Instead of asserting equality of density pairs 
at the sample level, we assess convergence at the distribution level: If 
$F_m(\rho)=|\psi|^2$ denotes the reconstruction of mass $m$ wave functions, 
then we search for the value of $m$ that leaves the \textsc{gravsphere} posterior over functions, $p(\rho)$, invariant under $F_m$, i.e. we demand:
$p(\rho) \overset{!}{=} p\left(F_m(\rho)\right)$.

The statistical framework to assert distribution equality if only samples, in our case sample functions, from both distributions are available, is a frequentist hypothesis test

\subsubsection{\label{sub:test} Details of the hypothesis test}

\noindent From a technical perspective, the idea underlying our hypothesis test is to 
(i) identify a metric $d:\mathcal{P}{(\mathcal{X})}\times\mathcal{P}{(\mathcal{X})}\to \mathbb{R}$ on the 
space of probability densities $\mathcal{P}(\mathcal{X})$ over square integrable functions defined 
on the spherical shell set by \textsc{gravsphere}'s validity region, 
$\mathcal{X} = L^2(\mathbb{S}(r_{1/2}/4, 4r_{1/2}))$, and (ii) use
an estimator for this distance measure as test statistic.

A suitable candidate for the metric, and the one we adopt in this work, is the
\emph{maximum mean discrepancy}
$\text{MMD}(p_x,p_y)=\sup_{f\in\mathcal{F}}|\mathbb{E}_{p_x}(f(x)) - \mathbb{E}_{p_y}(f(y))|$, over a, yet unspecified, class of functionals $\mathcal{F}=\{f:\mathcal X \to \mathbb{R}\}$.
A variety of classes is conceivable as discussed in
Ref.~\cite{Sriperumbudur2009}. For instance, if $X=\mathbb{R}$ and $\mathcal{F}=\{\mathbb{1}_{(-\infty,t]}:
t\in \mathbb{R}\}$, the set of indicator functions, one recovers the well-known 
Kolmogorov distance and the Kolmogorov-Smirnov statistic as its estimator.

Our specific choice is guided by two requirements:
On one hand, we demand that $\text{MMD}$ is indeed a metric, and thus in
particular that $\text{MMD}(p_x,p_y) = 0 \Rightarrow p_x=p_y$. On the other
hand, we require that an unbiased estimator for the $\text{MMD}$ is easily
computable. The authors of \cite{Gretton2012}
advocate to fix a kernel function $k:\mathcal{X}\times\mathcal{X} \to
\mathbb{R}$ and choose its reproducing kernel Hilbert space over
$\mathcal{X}$ as the function class over which MMD is computed. In accordance
with our preconditions on $\mathcal{F}$, one can show \citep{Gretton2012} that
with this choice the intractable optimisation step over $\mathcal{F}$ is
eliminated since 
$\text{MMD}_k(p_x,p_y)^2 = \mathbb{E}_{p_x}[k(x,x')]
-2\mathbb{E}_{p_x,p_y}[k(x,y)] +\mathbb{E}_{p_y}[k(y,y')]$ --- now void of any maximisation.
Moreover, if we fix $k_\gamma(x,y)=\exp\left(-\gamma||x-y||_\mathcal{X}^2\right)$ as our
kernel, Ref.~\cite{Wynne2020} proves that $\text{MMD}_{k_\gamma}(p_x,p_y)$ indeed constitutes a metric on
$\mathcal{P}(\mathcal{X})$. 

\begin{figure}
    \centering
    \includegraphics[width=\columnwidth]{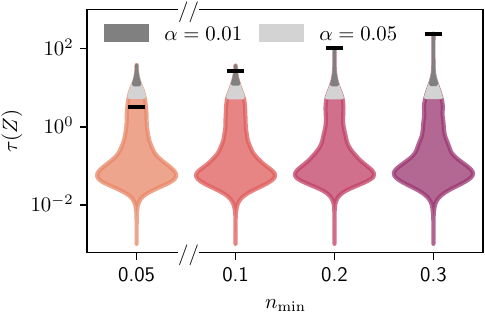}
    \caption{%
        Two-sample test run on the mock densities in Eq. \eqref{eq:mock}.
        This test emulates the fully fledged analysis of the main text by
        replacing the effective $m\to n(m)$ mapping with a uniform prior
        $U(n_\text{min},1)$ on the coredness parameter $n$. We find $Y_{n_\text{min}}$ to be
        indistinguishable from $X$ at significance $\alpha=0.05$ once
        $n_\text{min}=0.05$.
    }
    \label{fig:appendix1_hypothesis_test}
\end{figure}

If we let $Z$ be the pair $(X,Y)$, then a low variance, unbiased estimator
$\widehat{\text{MMD}}_{k_\gamma}(Z)$, is given in Refs.~\cite{Gretton2012, Biggs2023}.
To circumvent data-splitting, we fix the kernel bandwidth $\gamma$ as proposed
and tested in Ref.~\cite{Biggs2023}: instead of using only one heuristically deduced bandwidth, we deploy a selection of prior bandwidths $\gamma_n$ between $\min{d_{ij}}$ and $\max{d_{ij}}$ where the distance matrix element $d_{ij} = ||z_i - z_j||_\mathcal{X}$ and $z_i,z_j \in Z$. Each bandwidth produces a different estimate $\widehat{\text{MMD}}_{k_{\gamma_n}}$ and we fuse them into a consensus
statistic $\tau(Z)$ by taking their soft maximum via $\tau(Z) = \log\left[\sum_n\exp\left(\widehat{\text{MMD}}_{k_{\gamma_n}}(Z)\right)\right]$.

\subsubsection{A worked example: The coredness parameter}

\noindent It is useful to build some intuition for the interpretation of our
two-sample hypothesis test. Looking at the exemplary wave functions in Fig.
\ref{fig:gravsphere_posterior}, it is apparent that our reconstruction scheme
is effectively equivalent to a mapping, $m\to n(m)$, from boson mass $m$ to the
``coredness'' parameter $n$ of the \textsc{coreNFWtides} profile: 

For small $m$, all states, and in particular the cored $l=0$ modes, are broad. This leads to a best fit density with an extended core or, in terms of an effective \textsc{coreNFWtides} proxy, a density with a coredness close to unity. Large mass modes, by contrast, are characterised by a smaller spatial extent, realise cuspier best fits, and consequently effective core strengths of $n\simeq 0$. All other parameters should remain largely unchanged, as in practice we found that we can fit the NFW tail equally well within the hypothesis test limits across all considered boson masses. 

With this reasoning in mind, we may therefore gain
additional insight into the sensitivity of the result presented in the main text by running the two-sample test on:
\begin{equation}
    \label{eq:mock}
    \begin{split}
        X &= \{\rho_\text{cNFWt,i}(n_i) \mid n_i \sim U(0,1)\}_{i=1}^{5000}\;,\quad\\
        Y_{n_\text{min}} &= \{\rho_\text{cNFWt,i}(n_i) \mid n_i \sim U(n_\text{min},1)\}_{i=1}^{5000}\;,
    \end{split}
\end{equation}
with $0<n_\text{min}<1$ and all other parameters fixed to their original values
provided by \textsc{gravsphere}. The mock density set $Y_{n_\text{min}}$ takes on the role of our wave function ensemble.
Note that all other (hyper)parameters, such as cardinality of $X$, $Y_{n_\text{min}}$ or the number of resampling iterations, are identical to the main text.

Fig. \ref{fig:appendix1_hypothesis_test} shows the distribution of $\tau(Z)$
for increasing values of $n_\text{min}$ (decreasing boson mass). 
We find $n_\text{min} \lesssim 0.05$ to be
accepted at a significance $\alpha=0.05$. Put differently, a $5\%$ mismatch in the prior overlap of $n$ is already enough for the devised test to reject $\text{H}_0$, which we deem satisfactory. 
This simple example motivates the intuitive
interpretation of our hypothesis test in terms of detecting statistical discrepancies between sets of overlapping density profiles in the inner region of the dwarf.

\subsection{Robustness to Density Profile Modelling}

\begin{figure}
    \centering
    \includegraphics[width=\columnwidth, trim={0.1cm 0.1cm 0.1cm 0.1cm},clip]{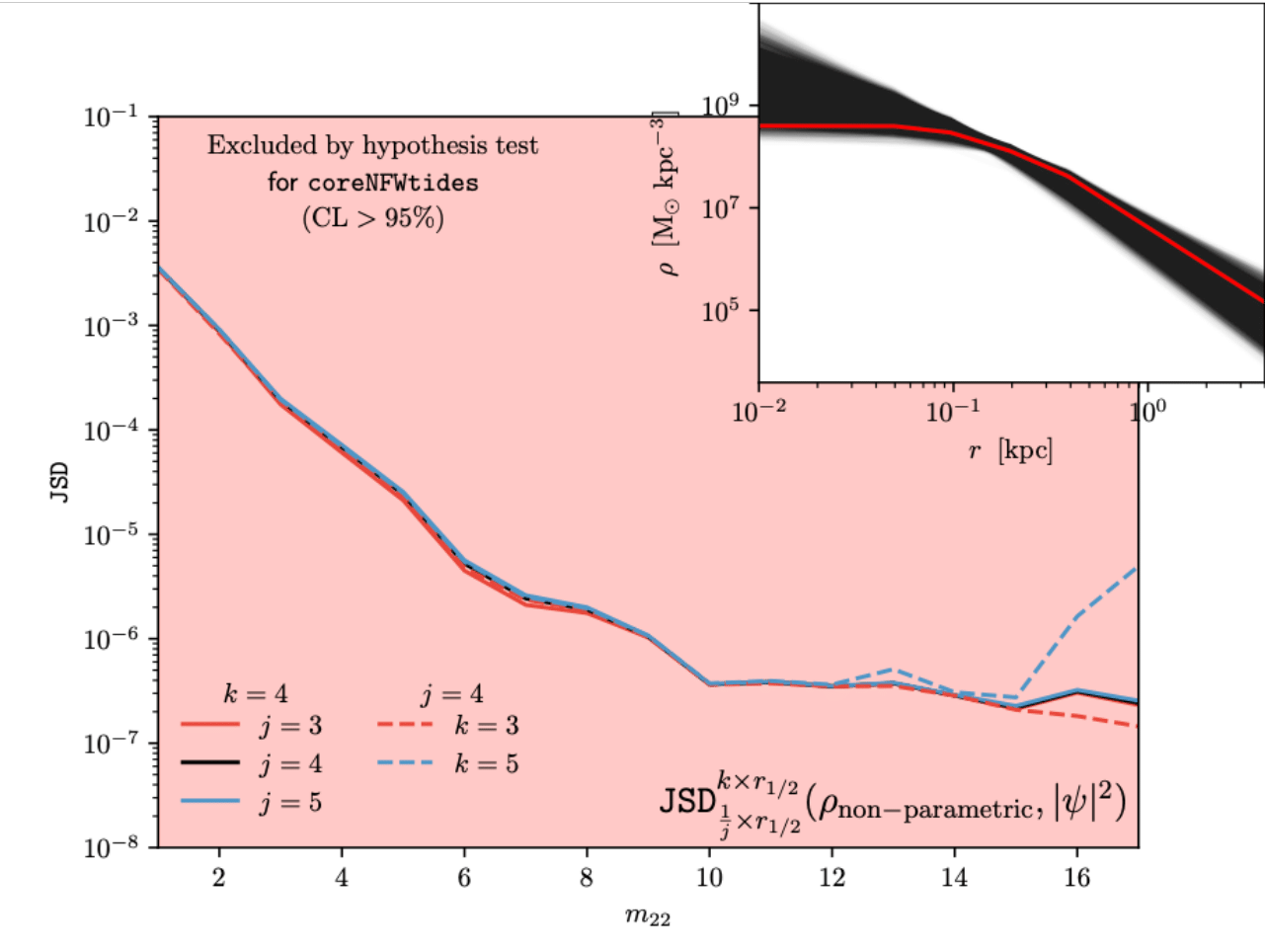}
    \includegraphics[width=\columnwidth, trim={0.1cm 0.1cm 0.1cm 0.1cm},clip]{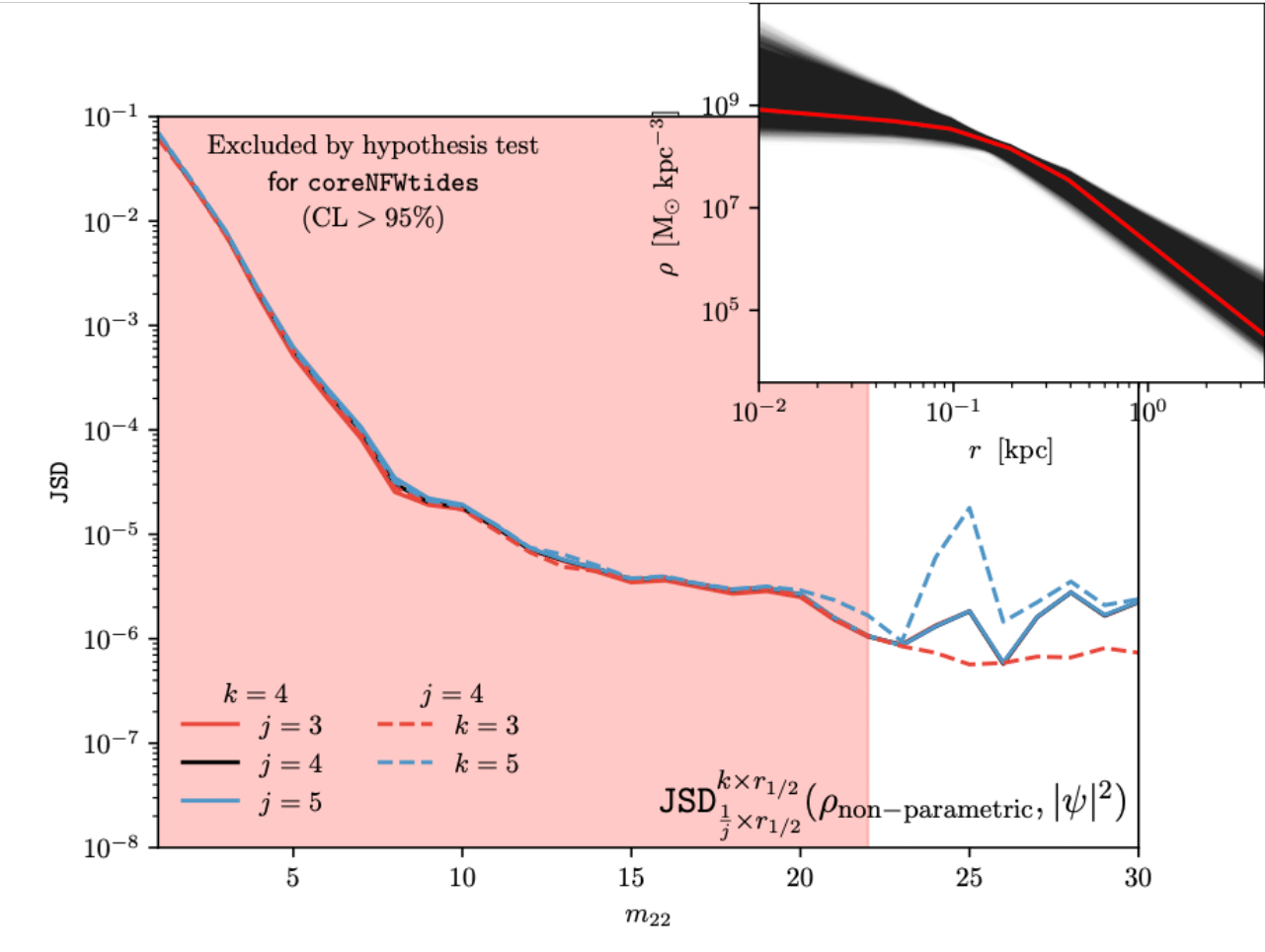}
    \includegraphics[width=\columnwidth, trim={0.1cm 0.1cm 0.1cm 0.1cm},clip]{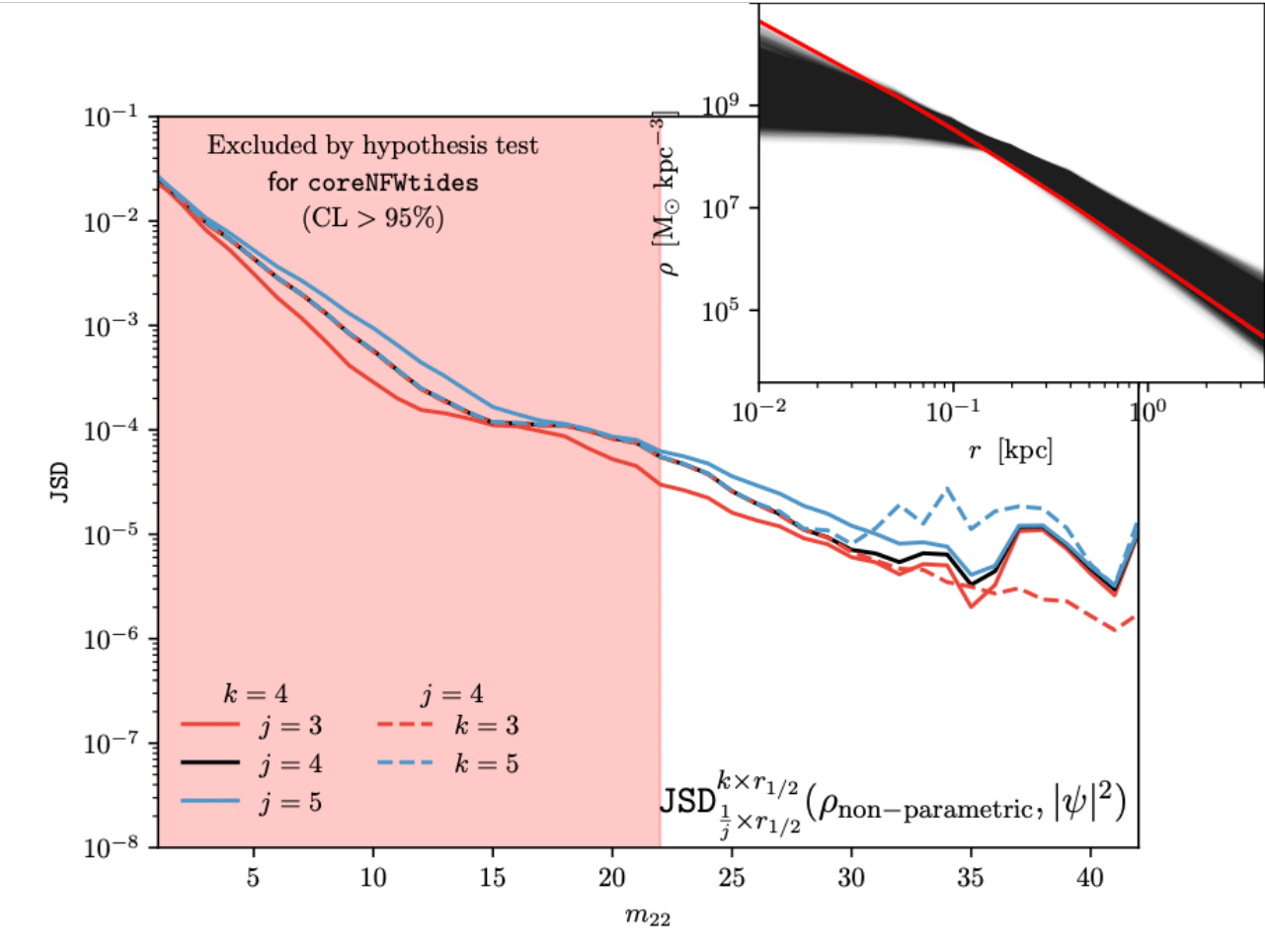}
    \caption{%
        Jensen-Shannon divergence (JSD) as a function of boson mass for a selection of density realistation of the free-form density population of Eq. \eqref{fig:non_parametric}. We choose $r_\mathrm{fit}=r_{200}$ as the wave function fit radius (to demonstrate robustness under changes in this parameter), and assess the robustness of the JSD under changes in the lower and upper cutoff. The case $k=j=4$ corresponds to the hypothesis limits in the main text. The area shaded in red depicts the excluded mass range according to the \texttt{coreNFWtides} ensemble. 
        }
        \label{fig:non_parametric}
\end{figure}

\noindent\textbf{Free-form density:}
We can build some intuition by re-doing the simplified, sample-based estimate of Fig. \ref{fig:simple_estimate} 
for a different parametrization of the posterior densities. 
We confine the discussion to the ``free-form/non-parametric`` model presented in \cite{Read2017}:
\begin{equation}
\frac{\rho_{\text{dm}}(r)}{\rho_0} =
\begin{cases}
 \left( \frac{r}{r_0} \right)^{-\gamma_0} & \hspace{-0.5em} r < r_0 \\[8pt]
 \prod\limits_{n=0}^{j < N_{\text{dm}}} \left( \frac{r_{n+1}}{r_n} \right)^{-\gamma_n} \left( \frac{r}{r_{j+1}} \right)^{-\gamma_{j+1}} & \hspace{-0.5em} r_j < r < r_{j+1}
\end{cases}
\end{equation}
with $N_{dm} = 5$ and fixed bins $r_{j=0 \dots (N_{\text{dm}} - 1)} = [0.25, 0.5, 1.0, 2.0, 4.0] \, r_{1/2}$.
Following the density draw selection strategy that led to Fig. \ref{fig:simple_estimate}, we find $m_{22}$ to be bracketed within $m_{22} \in [10, 32]$, see Fig. \ref{fig:non_parametric}. We thus expect a full ensemble estimate for this posterior to be similar to the result of the main text.

\vspace{0.6em}
\noindent\textbf{Spatial cutoff:}
While multiple spatial cutoffs exist in the analysis, two are of particular significance to our limit: the minimal/maximal radius setting the radial bin of the hypothesis test.
Other cutoffs, such as the fit radius, $r_\mathrm{fit}$, the maximal size of the classically allowed region, $r_\mathrm{lib}$, or the radius up to which the potential is computed,
$r_\mathrm{pot}$, are always set as to not interfere with the hypothesis test, i.e $r_\mathrm{max} \ll r_\mathrm{fit} \leq r_\mathrm{lib} \ll r_\mathrm{pot}$.

\begin{itemize}
    \item $r_\text{min}$: The impact of changing the lower bound will depend on the coredness of the profile to fit. Eigenmodes with vanishing angular momentum are cored. Decreasing $r_\text{min}$ on a cored or a slightly slanted profile will have none to negligible impact. The latter becomes plausible if we remind ourselves of the jacobian in spherical coordinates, further suppressing small radius deviations (thus making the chosen objective function conservative). By contrast, a reduced value of $r_\text{min}$ on a cuspy profile will drive the bound to higher masses. This can be seen in Fig. \ref{fig:non_parametric} where we change the lower bound of the JSD computation by $r_\text{min} = \frac{1}{j} \times r_{1/2}$. In the upper two cases, i.e. for a cored and slightly slanted inner profile all solid lines coincide, implying $r_\text{min}$ carries no importance in this regime of the posterior. The cuspy profile of the lower panel, by contrast, shows a stronger dependence on $r_\text{min}$ and, as intuitively expected, smaller values of $r_\text{min}$ translate to a larger JSD value.
    \item $r_\text{max}$: Increasing/decreasing $r_\text{max}$ has no impact on the convergence into the JSD plateau in Fig. \ref{fig:non_parametric} as the low mass end (and thus our limit) is driven by the behaviour in the inner region. Once convergence is achieved, increasing $r_\text{max}$ leads to stonger fluctuations inside the convergence plateau while decreasing $r_\text{max}$ dampens the oscillations. This effect is purely numerical and can be attributed to residual oscillations in the fitted wave function that exist in the halo outskirt where a (potentially very) large number of modes reside. Averaging over a larger number of random initial conditions for the optimiser can reduce this effect in principle. The point is that convergence (in the physically relevant core region) is achieved well before these oscillations matter. 
\end{itemize}

Extrapolating these intuitive, quantitative findings onto the impact for our ensemble-level limit, it is clear that $r_\text{max}$ is negligible, while a decreased $r_\text{min}$ can in principle yield a more competitive bound if the majority of posterior draws are cusped.
That said, lowering $r_\text{min}$ is only justified if the fidelity of the Jeans analysis
(and data resolution) allows for this. As stated in the main text, we chose $r_\text{min}=r_{1/2}/4$ for exactly this reason, since an application of \textsc{gravsphere} on LeoII-like mock data was able to recover the correct density up to this radius \cite{Collins2021}.

\vspace*{6pt}
\noindent\textbf{Sphericity:}
Fully relaxing the assumption of sphericity in our detailed analysis is highly non-trivial as it entails solving Schr\"odinger's equation in ellipsoidal symmetry. However, some intuition 
can be built by generalising the back of the envolope calculation in our introduction, which only relies on bulk properties, i.e. the virial radius $\sigma_r$ and velocity dispersion $\sigma_v$.

To estimate this, we deform a spherical dwarf into triaxial shape, such that isodensity surfaces obey:
\begin{equation}
    \frac{R^2}{c^2} = \frac{x^2}{a^2} + \frac{y^2}{b^2} + \frac{z^2}{c^2}\;,
\end{equation}
with ellipsoidal radius $R$ and minor/intermediate/major axis $a,b,c$. Assuming that the functional form of the density and potential do not change, such that after the deformation:
\begin{equation}
    \rho(r) \to \rho(R)\;\quad V(r)\to V(R)\;,
\end{equation}
it is straight forward to see that for fixed mass \cite{Smith2005}:
\begin{equation}
    R_{200} \equiv (\sigma_R)_\mathrm{triax} = \left(\frac{c^2}{ab}\right)^{1/3} (\sigma_r)_\mathrm{sph}\;.
\end{equation}

As in the main text of our paper, the virial theorem sets the velocity dispersion, which under triaxial conditions reads:
\begin{equation}
    (\sigma_v^2)_\mathrm{triax}=\frac{1}{2M_{\rm tot}}\left|\int\text{d}\bm x^3 \rho(\bm x)V(\bm x)\right| = \frac{ab}{c^2}(\sigma_v^2)_\mathrm{sph}\;.
\end{equation}
The uncertainty principle $\sigma_r\sigma_v \sim \frac{\hbar}{m}$ then suggests the mass bound for a triaxial configuration to be:
\begin{equation}
    m_\mathrm{triax} = \left(\frac{c^2}{ab}\right)^{1/6} m_\mathrm{sph}\;.
\end{equation}
Thus, we find a weak dependence of this  bound on the degree of non-sphericity (although not necessarily an improvement over the spherical case if the object is oblate $ab>c^2$). We expect this behaviour to translate to our analysis, as we implicitly apply the uncertainty principle on a per state basis when we fit against an envelope density profile. This is the case since the length scale that sets the spatial extent of each eigenmode is of the order of the per state uncertainty $\sigma^2_{r,{nl}} = \langle\psi_{nl}|r^2|\psi_{nl}\rangle - \langle\psi_{nl}|r|\psi_{nl}\rangle^2$.

\vspace*{6pt}
In the specific case of Leo II (and if $a=b$), \cite{Hayashi2020} reports a dark matter ellipticity of:
\begin{equation}
    Q_\text{LeoII} \equiv \frac{c}{a} = 1.08^{+0.61}_{-0.60},
\end{equation}
which (albeit somewhat unconstrained) is consistent with sphericity. For comparison, Draco, another candidate dwarf for our analysis, favours a more prolate configuration:
\begin{equation}
    Q_\text{Draco} = 1.39^{+0.40}_{-0.55}\;.
\end{equation}
More generally, it was found \cite{Read2017} that applying \textsc{gravsphere} on explicitly triaxial mock data, a comparison between the virial shape parameters computed from the posterior and the value of its data-based estimator will not coincide. Thus, independent data quality checks based on the virial shape parameters are able to pick up
systematic biases due to fitting non-spherical data with spherical mass models in \textsc{GravSphere} \cite{Read2017}.

\vspace{0.6em}
\noindent\textbf{Stationary Conditions:}
Dark matter composed of an ultralight bosonic species can induce dynamical heating of the stellar population due to the presence of a turbulent interference substructure or a randomly walking soliton.
The result of these effects is a spatial expansion of the stellar population and thus migration of the half-light radius. As noted by \cite{Levkov2018}, soliton formation (and consequently heating due to its random walk) is expected to be subdominant in dwarf-like environments if $m_{22}>20$.

For interference-induced heating, we rely on the minimal heating time scale derived in \cite{Bar2019}:
\begin{equation}
T_{\text{heat}}^{\text{min}}
\approx 7.5 \, \text{Gyr} \left( \frac{m}{2.3\times10^{-21} \, \text{eV}} \right)^{-1} \left( \frac{v}{10 \, \text{km s}^{-1}} \right)^{-2} \;.
\end{equation}
For comparison, the dynamical time scale of the stellar population is:
\begin{equation}
    T_{\text{dyn}} = \frac{R}{v} \approx 0.1 \, \text{Gyr} 
\left( \frac{R}{1 \, \text{kpc}} \right) 
\left( \frac{10 \, \text{km s}^{-1}}{v} \right)\;.
\end{equation}
This suggests heating to be a slow secular drift on the timescale of the tracer dynamics, and one should expect quasi-stationary conditions througout. The application of a stationary Jeans analysis (for the stars) is therefore permissible.
On the other hand, using the age of Leo II's main stellar population as proxy for its age, data suggests $T_\text{LeoII} = 9 \pm 1 \;\text{Gyr}$ \cite{Mighell1996}. Since $T_\text{LeoII}\gtrsim T_{\text{heat}}^{\text{min}}$, heating can in principle have led to an outwards migration of the stellar component. If so, this translates to a less spatially localised DM density posterior (cusps turn more core-like), and consequently a more conservative mass limit according to our analysis compared to the "true" value of the mass if such a secular drift would be taken into account. Our time scale-based argument is supported by fully-fledged simulations of a Leo II like dwarf, which due to presence of stellar heating, suggest $m_{22}\geq50$ \cite{Teodori2025}
\clearpage
\end{document}